\documentclass[twocolumn]{aastex631}
\usepackage{amsmath,amssymb}
\usepackage{tablefootnote}

\newcommand{\refeq}[1]{Eq.~(\ref{eq:#1})}          
          
\newcommand{\reffig}[1]{Figure~\ref{fig:#1}}          
\newcommand{\refsec}[1]{Section~\ref{sec:#1}}

\newcommand{\reftab}[1]{Table~\ref{tab:#1}}

\newcommand{\be}{\begin{equation}}
\newcommand{\ee}{\end{equation}}
\newcommand{\ba}{\begin{eqnarray}}
\newcommand{\ea}{\end{eqnarray}}

\newcommand{\nele}{n_{\rm e}}
\newcommand{\nH}{n_{\rm H}}
\newcommand{\Te}{T_{\rm e}}
\newcommand{\Ebv}{E({\rm B}-{\rm V})}
\newcommand{\nn}{\nonumber\\}

\newcommand{\ld}[1]{\textcolor{red}{[#1]}}

\shorttitle{Nitrogen-enriched nebula surrounding a super star cluster}
\shortauthors{Pascale et al.}

\begin{document}

\title{Nitrogen-enriched, highly-pressurized nebular clouds surrounding a super star cluster at cosmic noon}

\author[0000-0002-2282-8795]{Massimo Pascale}
\affiliation{Department of Astronomy, University of California, 501 Campbell Hall \#3411, Berkeley, CA 94720, USA}

\author[0000-0003-2091-8946]{Liang Dai}
\affiliation{Department of Physics, University of California, 366 Physics North MC 7300, Berkeley, CA. 94720, USA}

\author[0000-0003-1858-3892]{Christopher F. McKee}
\affiliation{Department of Astronomy, University of California, 501 Campbell Hall \#3411, Berkeley, CA 94720, USA}
\affiliation{Department of Physics, University of California, 366 Physics North MC 7300, Berkeley, CA. 94720, USA}

\author[0000-0002-6543-2993]{Benny T.-H. Tsang}
\affiliation{Department of Astronomy, University of California, 501 Campbell Hall \#3411, Berkeley, CA 94720, USA}
\affiliation{Theoretical Astrophysics Center, University of California, Berkeley, CA 94720, USA}



\begin{abstract}

Strong lensing offers a precious opportunity for studying the formation and early evolution of super star clusters that are rare in our cosmic backyard. The Sunburst Arc, a lensed Cosmic Noon galaxy, hosts a young super star cluster with escaping Lyman continuum radiation. Analyzing archival HST images and emission line data from VLT/MUSE and X-shooter, we construct a physical model for the cluster and its surrounding photoionized nebula. We confirm that the cluster is $\lesssim4\,$Myr old, is extremely massive $M_\star \sim 10^7\,M_\odot$ and yet has a central component as compact as several parsecs, and we find a {gas-phase metallicity} $Z=(0.22\pm0.03)\,Z_\odot$. The cluster is surrounded by $\gtrsim 10^5\,M_\odot$ of dense clouds that have been pressurized to $P\sim 10^9\,{\rm K}\,{\rm cm}^{-3}$ by perhaps stellar radiation at within ten parsecs. These should have large neutral columns $N_{\rm HI} > 10^{22.8}\,{\rm cm}^{-2}$ to survive rapid ejection by radiation pressure.
The clouds are likely dusty as they show gas-phase depletion of silicon, and may be conducive to secondary star formation if $N_{\rm HI} > 10^{24}\,{\rm cm}^{-2}$ or if they sink further toward the cluster center. Detecting strong {\rm N III]}$\lambda\lambda$1750,1752, we infer heavy nitrogen enrichment $\log({\rm N/O})=-0.21^{+0.10}_{-0.11}$. This requires efficiently retaining $\gtrsim 500\,M_\odot$ of nitrogen in the high-pressure clouds from massive stars heavier than $60\,M_\odot$ up to 4 Myr. We suggest a physical origin of the high-pressure clouds from partial or complete condensation of slow massive star ejecta, which may have important implication for the puzzle of multiple stellar populations in globular clusters.

\end{abstract}




\section{Introduction}
\label{sec:intro}

Massive star clusters represent the most prominent mode of star formation across cosmic history \citep{AB18}. In the hierarchical picture of star formation, massive clusters are considered to be the natural precursors of globular clusters \citep{Kruijssen19,KMBH2019ARAAreview}.
Understanding the internal dynamics of massive clusters during their formation and evolution episodes is crucial to properly answering many other important questions in astrophysics. Strong stellar feedback from young stars in massive clusters {in the form of winds} \citep{Silich04,Geen21,Lancaster21}, radiation pressure \citep{MQT10,Kim18,Kim19,Menon22a,Menon22b}, and supernova explosions \citep{WN15} are known to regulate star formation and drive outflows on galactic scale.
In addition, the extremely high stellar densities at the cores of massive clusters are conducive to stellar collisions and merging, a potential channel for seeding the formation of supermassive black holes \citep{PZ04,Davies11} and making gravitational wave sources \citep{Banerjee17,Banerjee18,Antonini19,Rodriguez19}. During the epoch of reionization, very young massive star clusters are capable of releasing escaping Lyman continuum radiation to the intergalactic medium (IGM) and might make an important contribution to cosmic reionization \citep{Vanzella2020IonizeIGM}.

With high-density birth environments being rare in our cosmic backyard, well-resolved dense star clusters of a few million years old are uncommon in the Local Group. A short list of examples include Westerlund 1~\citep{PortegiesZwart2010ARAAReview, Gennaro2017Westerlund1} and Westerlund 2~\citep{Ascenso2007Westerlund2}, the Arches~\citep{Espinoza2009ArchesClusterIMF, Harfst2010ArchesClusterInitCondition}, the Quintuplet~\citep{Figer1999QuintupletClusterMassiveStars}, and R136 in the Large Magellanic Cloud~\citep{Andersen2009R136IMF, Cignoni2015TarantulaTreasuryProject, Crowther2016R136a}. These have relatively small stellar masses $M_\star \sim 10^4\mbox{--}10^5\,M_\odot$. More massive systems $M_\star \sim 10^5\mbox{--}10^6\,M_\odot$, either exposed or obscured by gas, are more frequently found in nearby dwarf or massive starburst galaxies, such as in NGC 5253 \citep{Gorjian2001NGC5253ProtoGC, Turner2015NGC5253HighSFEfficiency}, NGC 1569 \citep{HoFilippenko1996NGC1569SSC}, M82 \citep{McCrady2007M82SSCs} and Henize 2-10 \citep{Chandar2003Hen210SuperStarClusters}. Recently, super star clusters with masses $M_\star = 10^6\mbox{--}10^7\,M_\odot$ were also observed in the nuclear region of NGC 1365 with the James Webb Space Telescope \citep{Whitmore2022SuperStarClustersNGC1365JWST}. Remarkably, extragalactic observations aided with strong gravitational lensing have revealed that very massive $M_\star \gtrsim 10^6\mbox{--}10^7\,M_\odot$ yet compact young systems appear commonplace in high-redshift star-forming galaxies~\citep{Vanzella2017protoGCs, Vanzella2022GLASSJWSTSSCA2744}. Strongly lensed systems hence offer a unique opportunity for studying clustered star formation at the highest densities.

In this work, we scrutinize a young massive star cluster found in the Sunburst Arc at $z_s=2.369$. The lensed galaxy was first discovered by \cite{Dahle2016SunburstDiscovery} behind the galaxy cluster PSZ1 G311.65-18.48 \citep{Planck2014SZClusterCatalog}. It is the brightest known gravitationally lensed starburst galaxy at Cosmic Noon. A lensed compact star-forming clump was found on the arc, which shows twelve different lensed avatars arising from an Einstein ring configuration disrupted by secondary lensing caustics cast by multiple cluster member galaxies \citep{RiveraThorsen2019Sci, Pignataro2021SunburstLensModel, Diego2022godzilla, Sharon2022SunburstLensModel}. 

The clump appears to be a super star cluster of about $3\,$Myr old \citep{Chisholm2019ExtragalacticMassiveStarPopulation}. Remarkably, optically thin channels must have been carved out through the surrounding interstellar medium (ISM), as both triple-peaked Ly$\alpha$ line radiation \citep{RiveraThorsen2017DirectLymanAlpha} and escaping Lyman continuum (LyC) radiation \citep{RiveraThorsen2019Sci} were detected from the super star cluster (hence we refer to it as the LyC cluster throughout this paper), with a high escape fraction $f_{\rm esc} = 40$--$90\%$ along the line of sight \citep{RiveraThorsen2019Sci}. High signal-to-noise ratio spectroscopy \citep{Mainali2022OutflowLyCescape} revealed evidences that the system drives outflowing nebular gas at $\sim 300\,{\rm km}\,{\rm s}^{-1}$, more dramatically so than in other Sunburst systems which do not allow LyC photons to escape. \cite{Mestric2023VMS} suggest that a few dozen $>100\,M_\odot$ very massive stars aggregate at the cluster center from which escaping LyC photons appear to originate, and can account for the observed broad {\rm He II}$\lambda$1640 emission.

Since the LyC cluster is a fantastic example of clustered massive star formation and evolution at Cosmic Noon and probably resembles the exposed LyC sources that have reionized the intergalactic medium (IGM) at $z =6$--$15$, much efforts have been made to understand its physical properties and the host galaxy. Rest-frame UV spectroscopy compared the LyC cluster of Sunburst to {\texttt Starburst99} spectral models \citep{Leitherer1999SB99} and inferred a cluster age $\sim 3\,$Myr from the broad {\rm He II} $\lambda1640$ emission, {a moderately sub-solar stellar metallicity} $Z=0.5$--$0.6\,Z_\odot$ from the presence of C IV $\lambda$1550 P Cygni stellar wind feature, and a moderate amount of dust reddening ${\rm E}(B-V)=0.15$ \citep{Chisholm2019ExtragalacticMassiveStarPopulation}.

\cite{Vanzella2020IonizeIGM} suggests that the LyC cluster spectrally resembles other known compact LyC leaking systems and should have an important contribution to cosmic reionizaton through holes created by stellar feedback. The cluster is shown to have strong emissions of (semi-)forbidden UV lines from O III, C III, Si III and Ne III \citep{Vanzella2020Tr}. Using multi-filter HST images to model the cluster SED and employing a detailed gravitational lensing model, \cite{Vanzella2022SunburstEfficiency} concluded that the cluster is impressively massive $M_\star \approx 10^7\,M_\odot$ yet compact $R_{\rm e}\simeq (8\pm 3)\,$pc. From emission line widths they derived a line-of-sight velocity dispersion $\simeq 40\,{\rm km}\,{\rm s}^{-1}$, which suggests that the cluster is gravitationally bound. The authors also suggested that the host galaxy has a high specific star formation rate sSFR$\gtrsim 10\,{\rm Gyr}^{-1}$.

In this work, we extract physical parameters describing the intriguing LyC cluster system and reliably estimate their uncertainties, by performing an independent spectral analysis using archival HST images and ground-based VLT/MUSE and VLT/X-shooter spectroscopy data. While the datasets we use (more details in \refsec{data}) are not new to what were used in the aforementioned studies, we present independent multi-filter photometry and emission line flux measurements. We then construct a physical model for the star cluster and the surrounding photoionization nebula based on the \texttt{BPASS} synthetic spectra \citep{Stanway2016BPASS, Eldridge2016BPASSbhmergers} for young stellar populations, and perform photoionization calculations using \texttt{Cloudy} \citep{Ferland2017c17}.

Our spectral modeling improves upon previous efforts in the following ways. First, we self-consistently account for the density dependence of nebular observables by imposing the more realistic isobaric condition \citep{Draine2011HIIRegionRadiationPressure, Kewley2019NebulaPressure} in \texttt{Cloudy} calculations. This is crucial as we find a highly pressurized component ($n_{\rm e}\sim 10^5\,{\rm cm}^{-3}$) in the photoionized nebula as deduced from {\rm C III]}$\lambda\lambda$1906,1908 and {\rm Si III]}$\lambda\lambda$1883,1892 line ratios.

Moreover, we account for nebular continuum radiation powered by radiative recombination in the photoionized gas, which is substantial for systems younger than $5\,$Myr \citep{Reines2010NebularContinuum, Byler2017NebularContinuum}. On one hand, the superposition of nebular continuum on top of the direct star light changes the apparent UV spectral slope and is crucial for our conclusion that dust reddening from the host ISM is small ${\rm E}(B-V) < 0.05$. On the other hand, the nebular continuum is proportional to the fraction of stellar ionizing radiation absorbed by the photoionized clouds; it enables us, when combined with nebular line strengths, to infer an isotropic LyC escape fraction $f_{\rm esc} = (35\pm 20)\%$ from the cluster vicinity. Furthermore, our inference of a higher line-of-sight escape fraction $(50\pm 20)\%$ is a hint that we are seeing exposed cluster stars through an optically thin channel.

Since photoionized clouds may have an anisotropic distribution around the cluster, our model is flexible to account for the spectral difference between illuminated cloud faces and shaded faces. This can be important for hydrogen Balmer lines due to optical depth effect from a large H I column, and for extinction of UV-to-optical continuum and lines by dust in the cloud interior. This probes the geometric distribution of photoionized clouds located on the near side versus on the far side of the stars relative to our viewing angle.

We furthermore use emission lines to measure element abundances for C, N, Ne and Si. From the detection of unusually strong {\rm N III]}$\lambda\lambda$1750,1752, we measure a remarkable N/O ratio enhanced by $\sim 12\mbox{--}19$ fold in the highly-pressurized clouds compared to what is expected for ISM at sub-solar metallicities. Knowledge of chemical anomalies informs us of the physical origin of the clouds, and provides clues to whether they might give birth to chemically self-enriched stellar population observed in evolved massive star clusters (i.e. globular clusters) \citep{Bastian2018ARAAGCMultiPop}. We also infer a depletion of gas-phase Si by $\sim 4\mbox{--}5$ fold, which indicates internal grain formation.  Curiously but consistent with many other observations of high-$z$ blue starburst systems, no strong dust reddening along the line of sight is seen from the host galaxy's large-scale ISM.

\begin{figure*}[t]
    \centering
    \includegraphics[width=\textwidth]{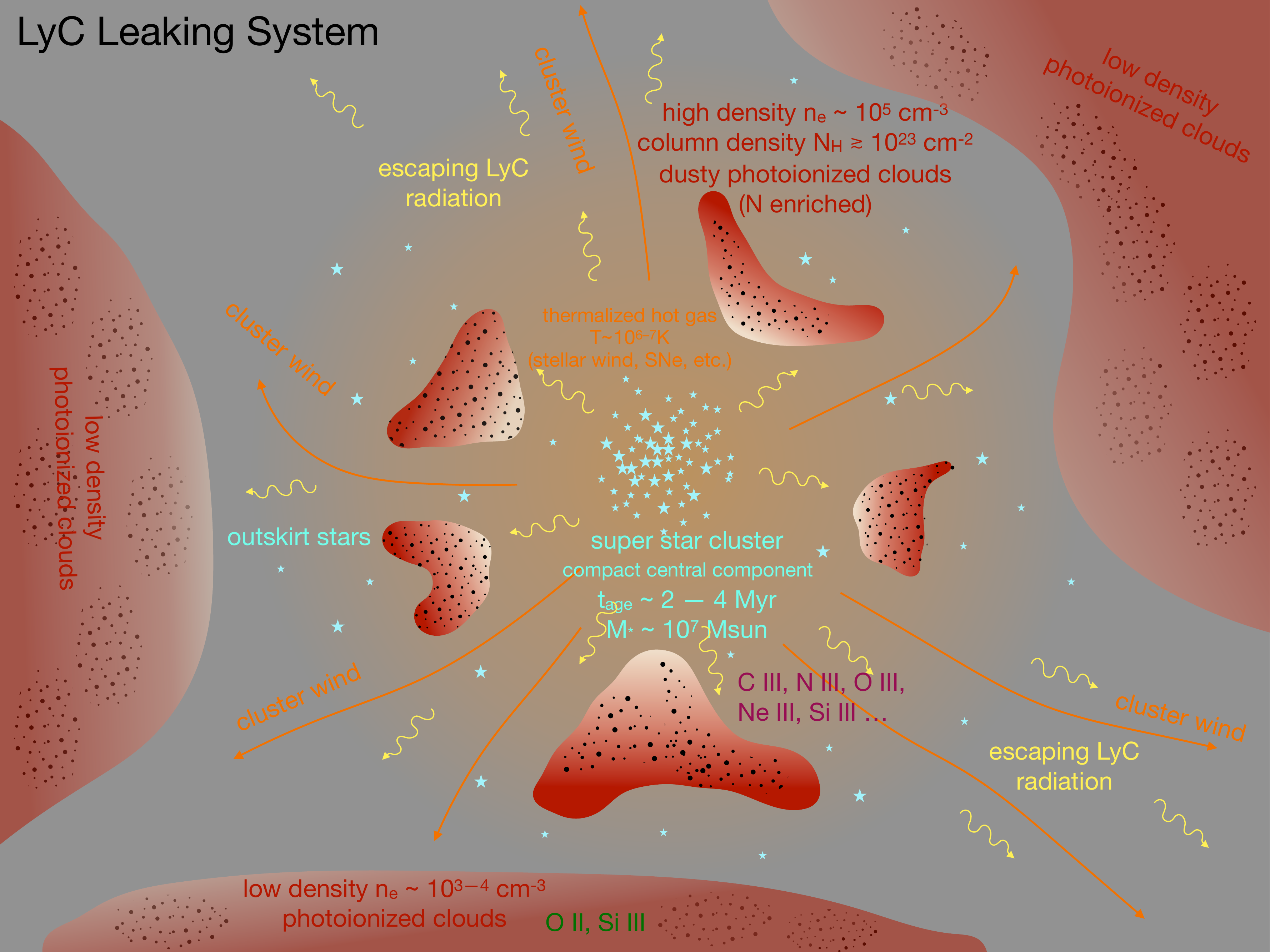}
    \caption{A cartoon illustrating a physical picture for the LyC leaking star cluster of Sunburst Arc derived from joint analysis of photometry and spectroscopy data. The central energy source, a super star cluster of $t_{\rm age} \lesssim4\,$Myr old, consists of a compact central component and an extended component of outskirt stars. Mass and kinetic energy output from both massive star winds and probably the first SN explosions thermalize and create a hot ($T\sim 10^6$--$10^7\,$K) cluster gas that blows out as a supersonic cluster wind. Pressed by stellar radiation (and possibly by the cluster wind), high-pressure ($\nele \sim 10^{5}\,{\rm cm}^{-3}$ and $P \sim 10^9$--$10^{10}\,{\rm K}\,{\rm cm}^{-3}$) clouds of large column density $N_{\rm H}\gtrsim 10^{23}\,{\rm cm}^{-2}$ linger close to the cluster at $\sim 5\mbox{--}10\,$pc, and are subject to a competition between outward pressure force and inward gravitational pull. Photoionized by the stars, these emit strong nebular lines from high-ionization ionic species, is chemically enriched by massive star ejecta, and is dense enough for dust formation. More clouds at lower densities ($\nele \sim 10^{3}$--$10^4\,{\rm cm}^{-3}$ and $P \sim 10^{7\mbox{--}8}\,{\rm K}\,{\rm cm}^{-3}$) exist at large distances, where nebular lines from low-ionization ionic species are excited. Photoionized clouds do not completely surround the star cluster, and a fraction of the LyC radiation escapes the system via low-density channels.}
    \label{fig:LyC_cartoon}
\end{figure*}

Based on our analysis, we depict the following physical picture for the star cluster and its surroundings as illustrated by \reffig{LyC_cartoon}. A young cluster of $t_{\rm age} \lesssim4\,$Myr and sub-solar {gas-phase} metallicity $Z\approx 0.2$--$0.3\,Z_\odot$ consists of a compact central component and a subdominant extended component. Intense stellar radiation pressurizes nearby photoionized gas to $P = {\rm a\,\,few}\times 10^9\,{\rm K}\,{\rm cm}^{-3}$ at several parsecs, while dust grains form inside these dense clouds. The clouds are significantly enriched with nitrogen and hence an order-unity mass fraction must have originated from massive star ejecta \citep{Roy2022PreSNwindsNitrogen}, perhaps through a mechanism of slow mass loss. Under irradiation by stars, these clouds glow in strong emission lines, processing $\sim 40\%$ of the cluster's LyC output. These clouds might linger for a time much longer than the free-fall time in the cluster's gravitational potential well, probably because they have large enough column densities $N_{\rm H} > 10^{22.8}\,{\rm cm}^{-2}$ to counter ejection by radiation pressure. Some other $\sim 20\%$ of the cluster's LyC output might be processed by lower-pressure photoionized clouds further from the cluster center, but those do not appear to have significant nitrogen enrichment. The remaining $\sim 40\%$ of the LyC radiation, together with the hot cluster gas, stream out of the system without obstruction through preferentially optically thin, low density channels.

The rest of the paper is organized as the following. In \refsec{data} we summarize the archival imaging and spectroscopy data used in our analysis. In \refsec{phot} and \refsec{spec}, we provide details on how HST filter magnitudes and emission line fluxes are measured. In \refsec{specmodel}, we employ stellar population synthesis and photoionization calculations to construct a grid of spectral models for the composite emission from the star cluster and its surrounding nebula. We then fit our spectral models to data, and present Bayesian parameter inference results in \refsec{results}. We give physical interpretation of the results in \refsec{disc}, and give concluding remarks in \refsec{concl}.

\section{Data}
\label{sec:data}

In this section, we summarize the archival data used in this work: HST imagining data, VLT/MUSE IFU spectroscopy, and VLT/X-shooter slit spectroscopy.

\subsection{{\textit HST} Imaging}

Observations for PSZ1 G311.65-18.48 were obtained across multiple programs between 2018 and 2020 (P.I. Dahle Program ID 15101, P.I. Gladders Program ID 15949, P.I. Bayliss Program ID 15377). In this work, we make use of observations publicly available on the MAST archive for 15 filters: WFC3/UVIS F390W, WFC3/UVIS F410M, WFC3/UVIS F555W, WFC3/UVIS F606W, ACS/WFC F814W, WFC3/UVIS F098M, WFC3/IR F105W, WFC3/IR F125W, WFC3/IR F126N, WFC3/IR F128N, WFC3/IR F140W, WFC3/IR F153M, WFC3/IR F160W, WFC3/IR F164N, WFC3/F167N. All UVIS+WFC images were aligned to a common pixel scale of 30 mas/pixel, while all IR images were aligned to a pixel scale of 60 mas/pixel using the astropy \texttt{reproject} package in combination with \texttt{astroalign} \citep{Beroiz2020}.

\subsection{VLT/MUSE IFU Spectroscopy}

We use archived and reduced MUSE IFU science datacubes directly accessed from the ESO archive\footnote{\url{http://archive.eso.org/scienceportal/home}}. Two datacubes are used, which cover the observed wavelength range $475$--$935\,$nm at $R\approx 2600$--$3000$ and were calibrated and reduced with the standard reduction pipeline: one from observations conducted on May 13 and Aug 24 in 2016 (Program 107.22SK; PI: E. Vanzella; total exposure $4449\,$s), and another from observation on Aug 10, 2021 (Program 297.A-5012; PI: N. Aghanim; total exposure $2808\,$s). All MUSE IFU observations were carried out in the Narrow Field Mode under a typical seeing condition FWHM $= 0.5$--$0.6\arcsec$, achieving a sky resolution $1.4$--$2\arcsec$.

\subsection{VLT/X-shooter Slit Spectroscopy}

We use ESO archived science data products of X-shooter slit spectroscopy reduced with the standard reduction pipeline. Data were collected from Program 0103.A-0688 (PI: E. Vanzella). Exposures were obtained under typical seeing conditions ${\rm FWHM}\sim 0.5\mbox{--}0.6\arcsec$. Slit spectra covering observed 994--2479 nm (R$\approx$8000) are used for extracting fluxes for rest-frame NUV and optical emission lines, while additional spectra covering observed 534--1020 nm (R$\approx$11000) are used for cross-checking strong FUV emission lines. For accessing spatially resolved spectral information, we carry out flux calibration of the 2D spectra using the flux-calibrated and flux-uncalibrated 1D spectra, and then extract fluxes at the location of the target objects.

\section{Photometry}
\label{sec:phot}

We measure the PSF of each HST image using the \texttt{photutils} package by stacking isolated, unsaturated stars across the image. An initial catalog of stars is generated using \texttt{DAOStarFinder} following \cite{Stetson1987}, and then each star is vetted by eye to be isolated and unsaturated, resulting in a final catalog of dozens of stars. The final star list is extracted into centered cutouts which are normalized and median-subtracted to mitigate sky background, and the cutouts are stacked into an oversampled PSF using the \texttt{EPSFBuilder} function. PSFs in the ACS filters were oversampled by a factor of 8, while WFC3 were only oversampled by a factor of 4 due to the lower native resolution. We found using stars throughout the entire image rather than those local to the LyC image produced more accurate PSFs, owing to {the nearly a factor of 10} greater number of stars available.

\subsection{Image 9}

There are a total of 12 lensed images of the LyC cluster available for analysis \citep{RiveraThorsen2019Sci}. Those have different magnification factors and are contaminated to various degrees by other sources on the arc.  We perform PSF photometry on Image 9 (following designation of \cite{RiveraThorsen2019Sci}) using \texttt{Galfit} \citep{Peng2002}. We choose Image 9 due to its separation from other bright LyC images and neighboring galaxies, as well as its moderate brightness relative to other LyC images. The brightest lensed images, which are the most magnified, are generally poorly approximated by a PSF (Image 10 is an example), whereas the fainter images have lower SNRs - Image 9 offers a compromise between these two effects. For Image 9, we simultaneously model both the LyC cluster and the neighboring lensed knots as point-like objects in \texttt{Galfit}, and approximate the underlying diffuse arc background as having a flat surface brightness. To mitigate large-scale variations in the diffuse background, we first mask out fainter pixels with brightness $<N\sigma$ to fit only the lensed arc, where $\sigma$ is the noise standard deviation of local sky background and $N$ is a factor of order dozens chosen by eye on an image-to-image basis. We found that different choices for $N$ caused small changes in the measured flux, and adjust the errors accordingly. We assign minimum errors of 0.05 and 0.1 magnitudes in the WFC3/UVIS+WFC/ACS and WFC3/IR bands respectively to broadly account for systematics. Our photometry results are reported in \reftab{photometry}.

\subsection{Image 10}

Because of the higher magnification factor, an extended continuum component of the LyC cluster appears resolved in Image 10 \citep{Vanzella2020IonizeIGM, Mainali2022OutflowLyCescape}, in addition to a bright central component. While the central component appears well-approximated by the PSF, the extended component extends $\sim 0.4\arcsec$ along the direction of the arc. The magnification in the radial direction, $\mu_r$, is expected to be of order unity (see \S 7.1), such that the extended component is only resolved in the tangential direction. Hence we choose a 1D Gaussian convolved with the PSF to approximate the light profile of the extended component. Retaining the same masking approach for the arc background used for image 9, we spatially fit a 1D Gaussian for the extended component, an unresolved central component, and  the arc background simultaneously using \texttt{PyMultinest} \cite{Feroz2009,Buchner2014}. This allows for separate photometric measurements for both the central and extended components, which we perform only on the WFC3/UVIS and ACS/WFC bands due to their sharper PSF's when compared to the IR bands. This is done for F555W, F606W and F814W, for which spatial resolution is good and continuum light is unaffected by Lyman alpha radiation transfer. Similar to the Image 9 photometry, we assign minimum errors of 0.05 and 0.1 magnitudes in the WFC3/UVIS+WFC/ACS and WFC3/IR bands respectively.

\subsection{Magnification ratio between Images 10 and 9}
\label{sec:muratio}

Comparing the flux of Image 10 (sum of central and extended components) to that of the unresolved Image 9, we derive that the magnification ratio is $\mu_{10}/\mu_{9} = 4.0-4.1$, consistent between F555W, F606W and F814W. However, we measure a magnification ratio $\mu_{10}/\mu_{9} = 3.4\pm 0.1$ in the F275W filter which probes escaping LyC radiation. Interestingly, the extended component is undetected in Image 10 {for the F275W filter which is sensitive only to LyC photons} \citep{Mestric2023VMS,Mainali2022OutflowLyCescape}. The discrepancy in the magnification ratio suggests that the extended component has a higher magnification than the central component. Such magnification variation across a small angular scale $\gtrsim 0.5\arcsec$ {are not well-described by the large-scale cluster lens}, and could be caused by a nearby invisible dark matter subhalo.
{While lens modeling predicts that Image 9 and Image 10 should have similar magnifications \citep{Pignataro2021SunburstLensModel, Sharon2022SunburstLensModel}, photometric measurements from \cite{Vanzella2022SunburstEfficiency} also confirmed a significant flux ratio of $\gtrsim2$, albeit lower than those found in this work.}
This might be a hint for spatial fine structures in the lens model near Images 9 and 10. In this work, we do not delve into investigating the physical cause of this effect, but use separate magnification ratios as priors for subsequent analyses: $\mu_{10}/\mu_{9} = 3.4 \pm 0.1$ for the central component, and $\mu_{10}/\mu_{9}= 5.6 \pm 0.5$ for the extended component. We shall assume that the unresolved Image 9 has a single magnification factor $\mu_9$ that applies to both components.

\begin{table*}
	\centering
	\caption{Photometry for lensed Image 9 and Image 10 of the LyC cluster.}
	\label{tab:photometry}
	\begin{tabular}{rcccl}
		\hline
		\hline
		Filter & Img. 9 & Img. 10 central & Img. 10 extended & Comment \\
		\hline
        F275W & $26.62 \pm 0.05$ & $25.27 \pm 0.05$ & $>28.6$ & Rest-frame LyC; no extended component and not included in analysis \\
		F390W & $23.03 \pm 0.05$ & $22.09\pm 0.05$ & $22.74\pm 0.05$ & Affected by Ly$\alpha$ and damping wings; not included in analysis \\
		F410M & $22.88 \pm 0.05$ & $22.04\pm 0.05$ & $22.29\pm 0.05$ & Strongly affected by Ly$\alpha$ and damping wings; not included in analysis \\
            \hline
		F555W & $22.95 \pm 0.05$ & $21.91 \pm 0.05$ & $22.56 \pm 0.05$& \\
		F606W & $22.91 \pm 0.05$ & $21.92 \pm 0.05$ & $22.46 \pm 0.05$ & \\
		F814W & $23.10 \pm 0.05$ & $22.14 \pm 0.05$ & $22.57 \pm 0.05$ &  \\
		F098M & $23.10 \pm 0.07$ & --- & --- & \\
		F105W & $23.01 \pm 0.10$ & --- & --- & \\
        F125W & $23.05 \pm 0.10$ & --- & --- & \\
		F126N & $22.50 \pm 0.20$ & --- & --- & Strongly enhanced by [O II]$\lambda\lambda$3727,3729 \\
        F128N & $23.01 \pm 0.10$ & --- & --- & No strong lines \\
		F140W & $23.01 \pm 0.10$ & --- & --- & Enhanced by multiple NUV lines. \\
		F153M & $23.57 \pm 0.10$ & --- & --- & No strong lines \\
		F160W & $22.57 \pm 0.10$ & --- & --- & Enhanced by mainly H$\beta$ and [O III]$\lambda\lambda$4959,5007 \\
		F164N & $22.55 \pm 0.10$ & --- & --- & Strongly enhanced by H$\beta$, not included in analysis \\
        F167N & $21.61 \pm 0.10$ & --- & --- & Strongly enhanced by [O III]$\lambda$4959 \\ 
		\hline
        \hline
	\end{tabular}
\end{table*}

\section{Spectroscopy}
\label{sec:spec}

\subsection{VLT/MUSE}

In order to constrain physical properties of the LyC cluster and its surrounding nebula using line ratio information, we extract rest-frame UV emission line fluxes from VLT/MUSE IFU data. Limited by seeing, MUSE IFU imaging has poor spatial resolution compared to HST images, and it is difficult to separate the lensed images of the LyC cluster from nearby sources on the arc. Since Image 10 is the brightest among the 12 lensed images of the cluster \citep{RiveraThorsen2019Sci}, we apply aperture photometry to it to robustly extract emission line fluxes. Using any of the other 11 fainter lensed images degrades the accuracy of line flux measurements due to reduced SNRs.

At a pixel scale $0.2\arcsec$, we extract fluxes from a $4\times 4$ aperture placed on top of Image 10. The aperture size is suitably chosen so that line contamination from nearby (unresolved) sources on the arc is small while a significant fraction (but not entirety) of the LyC cluster's flux is included. Sky background is then subtracted by extracting fluxes from an identical aperture placed at an empty location off the arc. Aperture correction is estimated by placing the same aperture at a nearby isolated star and comparing the extracted flux to that extracted from a sufficiently large aperture that includes all flux of the star. Within the MUSE wavelength coverage, the aperture correction factor varies from $\sim 2.6$ to $\sim 2.3$ with a mild wavelength dependence.

A number of rest-frame FUV lines are robustly detected \citep{Vanzella2020Tr}, all appearing to have the same systemic redshift. We fit each line to a Gaussian line profile on top of the local continuum, and flux uncertainty is estimated based on local flux fluctuations. Two MUSE IFU datacubes are used, one from combined exposures on May 13, 2016 and Aug 24, 2016, and another from observation on Aug 10, 2021. We found that all detected lines have fluxes consistent between the two datacubes (corresponding to an observed baseline of $~5$ years, or $~1.5$ years in the source frame) within the noise level, thus finding no evidence of spectral variability. We then combine measurements from both datacubes using inverse variance weighting. We report isotropic line luminosities multiplied by magnification in \reftab{MUSE_lines}, which are aperture corrected but not dust reddening corrected.

\begin{table}
	\centering
	\caption{Inferred isotropic luminosities of nebular emission lines detected from VLT/MUSE IFU data for Image 10 of the LyC cluster. The magnified isotropic line luminosity is calculated for a luminosity distance of $d=19.566\,$Gpc (corresponding to the source redshift $z_s=2.369$), and is uncorrected for MW dust reddening.}
	\label{tab:MUSE_lines}
	\begin{tabular}{rc}
		\hline
		\hline
		Emission line & Luminosity $\mu\,L\,[10^{41}\,{\rm erg}\,{\rm s}^{-1}]$ \\
		\hline
        {\rm [O III]}$\lambda$1660 & $5.8\pm 2.2$ \\
        {\rm [O III]}$\lambda$1666 & $17.0\pm 1.6$ \\
        {\rm N III]}$\lambda$1750 & $13.7\pm 2.7$  \\
        {\rm N III]}$\lambda$1752 & $6.8\pm 1.6$  \\
        {\rm [Si III]}$\lambda$1883 & $4.8\pm 1.6$ \\
        {\rm Si III]}$\lambda$1892 &  $8.5\pm 1.0$\\
        {\rm [C III]}$\lambda$1906 & $22.9\pm 1.6$\\
        {\rm C III]}$\lambda$1908 & $45.7\pm 3.7$\\
        {\rm [O II]}$\lambda$2471 & $5.1\pm 1.0$ \\
		\hline
        \hline
	\end{tabular}
\end{table}

\subsection{VLT/X-shooter}

Additional rest-frame NUV and optical lines have been detected from slit spectroscopy with VLT/X-shooter \citep{Vanzella2020Tr}. In the 2D spectra (for a $0.8\arcsec \times 11\arcsec$ slit), we apply for a wavelength dependent flux calibration factor calculated as the ratio between the 1D spectrum and the reduced 1D spectrum. Flux is extracted from a $0.8\arcsec\times 1.27\arcsec$ segment on the slit centered at Image 10. For each line, we fit a Gaussian line profile on top of the continuum. Since it is difficult to empirically determine the slit correction factor based on available data, we estimate a theoretical flux correction factor $=1.2$ by modeling the effect of seeing as a Moffat PSF profile \citep{Trujillo2001MoffatPSF} calculated at the mean seeing FWHM=$0.54\arcsec$, and approximate it as wavelength independent. Inferred line luminosities multiplying magnification but without dust reddening correction are reported in \reftab{Xshooter_lines}.

\begin{table}
	\centering
	\caption{Inferred isotropic luminosities of nebular emission lines detected from VLT/X-shooter for Image 10 of the LyC cluster. Magnified line luminosities are reported, without correcting for dust reddening. We have applied a wavelength independent PSF correction factor $=1.2$ using a theoretical Moffat PSF model with $\beta=4.765$.}
	\label{tab:Xshooter_lines}
	\begin{tabular}{rc}
		\hline
		\hline
		Emission line & Luminosity $\mu\,L\,[10^{41}\,{\rm erg}\,{\rm s}^{-1}]$ \\
		\hline
        {\rm [O II]}$\lambda$3726 & $36\pm 7$ \\
        {\rm [O II]}$\lambda$3729 & $18\pm 11$ \\
        {\rm [Ne III]}$\lambda$3869 & $47\pm 5$ \\
        {\rm [Ne III]}$\lambda$3967 & $13\pm 4$ \\
        H$\beta$ &  $94\pm 32$\\
        {\rm [O III]}$\lambda$4959 & $295\pm 36$\\
        {\rm [O III]}$\lambda$5007 & $864\pm 86$\\
        {\rm [N II]}$\lambda$6548 & $2\pm 11$ \\
        H$\alpha$ & $335\pm 108$ \\
        {\rm [N II]}$\lambda$6583 & $0\pm 11$ \\
		\hline
        \hline
	\end{tabular}
\end{table}

\section{Spectral Modeling}
\label{sec:specmodel}

We jointly fit HST photometry as well as line fluxes and ratios from MUSE and X-shooter to spectral models that incorporate both direct stellar radiation and nebula reprocessed (continuum and line) radiation from photoionized clouds. In this way, we can infer physical parameters that characterize the star cluster and the surrounding nebula. Below we discuss how we construct the spectral models.

\subsection{Star Cluster SED}
\label{sec:stars}

We use incident stellar spectra from the population synthesis model BPASS (v2.0) \citep{Stanway2016BPASS, Eldridge2016BPASSbhmergers} as the input for photoionization calculations. Our default choice is that the LyC cluster can be modeled as an instant burst of star formation, and we include output radiation from binaries. A stellar initial mass range $0.1 < m/M_\odot < 300$ is used with an IMF slope $\xi(m) \propto m^{-1.3}$ for $m < 0.5\,M_\odot$ and a slope $\xi(m) \propto m^{-2.35}$ for $m > 0.5\,M_\odot$. We later refer to this as the standard IMF. We choose a high mass cutoff $m < 300\,M_\odot$ because the system is very massive $M \gtrsim 10^6\,M_\odot$ \citep{Vanzella2020IonizeIGM}. The incident stellar spectrum is parameterized by cluster age $t_{\rm age}$ and metallicity $Z$. While BPASS uses a solar metallicity value $Z_\odot=0.02$ to which early stellar atmosphere models were calibrated to, the solar nebular metal abundance we shall use later for photoionization calculation with \texttt{Cloudy} is $Z_\odot=0.014$, which is the preferred value according to \cite{Asplund2021SolarAbund2020}. 

\subsection{Photoionization Calculations}
\label{sec:nebula}

Given the incident stellar radiation, we use the photoionization code \texttt{Cloudy} (v17) \citep{Ferland2017c17} to compute nebular continuum and line emissions, as well as the attenuation of star light through the gas. For all except three elements, we set their nebular abundances to be a factor $Z/Z_\odot$ of \texttt{Cloudy}'s default solar abundance (Table 7.1 of \texttt{Cloudy} manual \texttt{HAZY 1}).
For our purposes, those are consistent with the more recent compilation in \cite{Asplund2021SolarAbund2020}, and they correspond to $Z_\odot=0.014$. We will use $Z_\odot=0.014$ when reporting gas metallicity in units of solar metallicity.

Baseline abundances for helium, carbon and nitrogen are adjusted as a function of $Z/Z_\odot$ to account for nucleosynthesis \citep{Dopita2006SEDstarburst}. For helium, we correct for primary nucleosynthesis in addition to primordial abundance \citep{RussellDopita1992abunMagellanic, Pagel1992HeAbunHII}, He/H=$0.0737 + 0.024\,(Z/Z_\odot)$. For carbon and nitrogen, we adopt the enrichment model of \cite{Dopita2006SEDstarburst}, using C/H=$6\times 10^{-5}\,(Z/Z_\odot) + 2\times 10^{-4}\,(Z/Z_\odot)^2$, and N/H=$1.1\times10^{-5}\,(Z/Z_\odot) + 4.9\times 10^{-5}\,(Z/Z_\odot)^2$, respectively. Through our analysis, we find significant depletion of gas phase silicon, which indicates grain formation. Hence, we reduce the baseline nebular silicon abundance by a fiducial factor of six in our models. Additionally, we find a strikingly elevated nitrogen abundance $\log({\rm N/O}) \simeq -0.21$ which is more dramatic than local nitrogen enrichment in H II regions around low metallicity young starbursts \citep{Izotov2006logZemislingalaxies}. We therefore apply a fiducial nitrogen enhancement by a factor of four on top of the \cite{Dopita2006SEDstarburst} baseline. Later when fitting emission lines, we will introduce abundance rescaling factors as free parameters relative to these baseline abundances, for C, N, Si and Ne; we set the baseline abundances to prevent big errors in gas cooling calculations in \texttt{Cloudy}.

We compute isobaric models, which are more realistic than constant density models \citep{Kewley2019NebulaPressure}. To this end, we compute a grid for $6 < \log P [{\rm K}\,{\rm cm}^{-3}] < 12$, where $P$ is the sum of gas pressure and radiation pressure, and is kept constant throughout the cloud. 

In all runs, the cloud geometry is that of a plane parallel slab, {calculated up to a fixed neutral column density $N_{\rm HI}=10^{22.5}\,{\rm cm}^{-2}$}. This is very close to a sufficient value to prevent the (dusty) clouds from being ejected out of the cluster's gravitational potential under radiation pressure~\citep{KMBH2019ARAAreview}. 

We compute for a grid of the standard ionization parameter $U$ by varying an artificial cloud distance to the stars. While our results suggest that external ISM dust reddening is minimal toward the LyC cluster, gas-phase silicon depletion combined with the high inferred gas density indicates that dust grains should have efficiently formed in the cloud interior. We therefore set a dust-to-gas ratio that is rescaled by $Z/Z_\odot$ from the standard value 1\% in the solar-metallicity ISM gas. 

Furthermore, we set a fixed velocity dispersion $3\,{\rm km}\,{\rm s}^{-1}$ for gas micro turbulence, which has unimportant effects on the nebular spectrum.

\subsection{Viewing Geometry}
\label{sec:viewing}

\begin{figure}[t]
    \centering
    \includegraphics[scale=0.25]{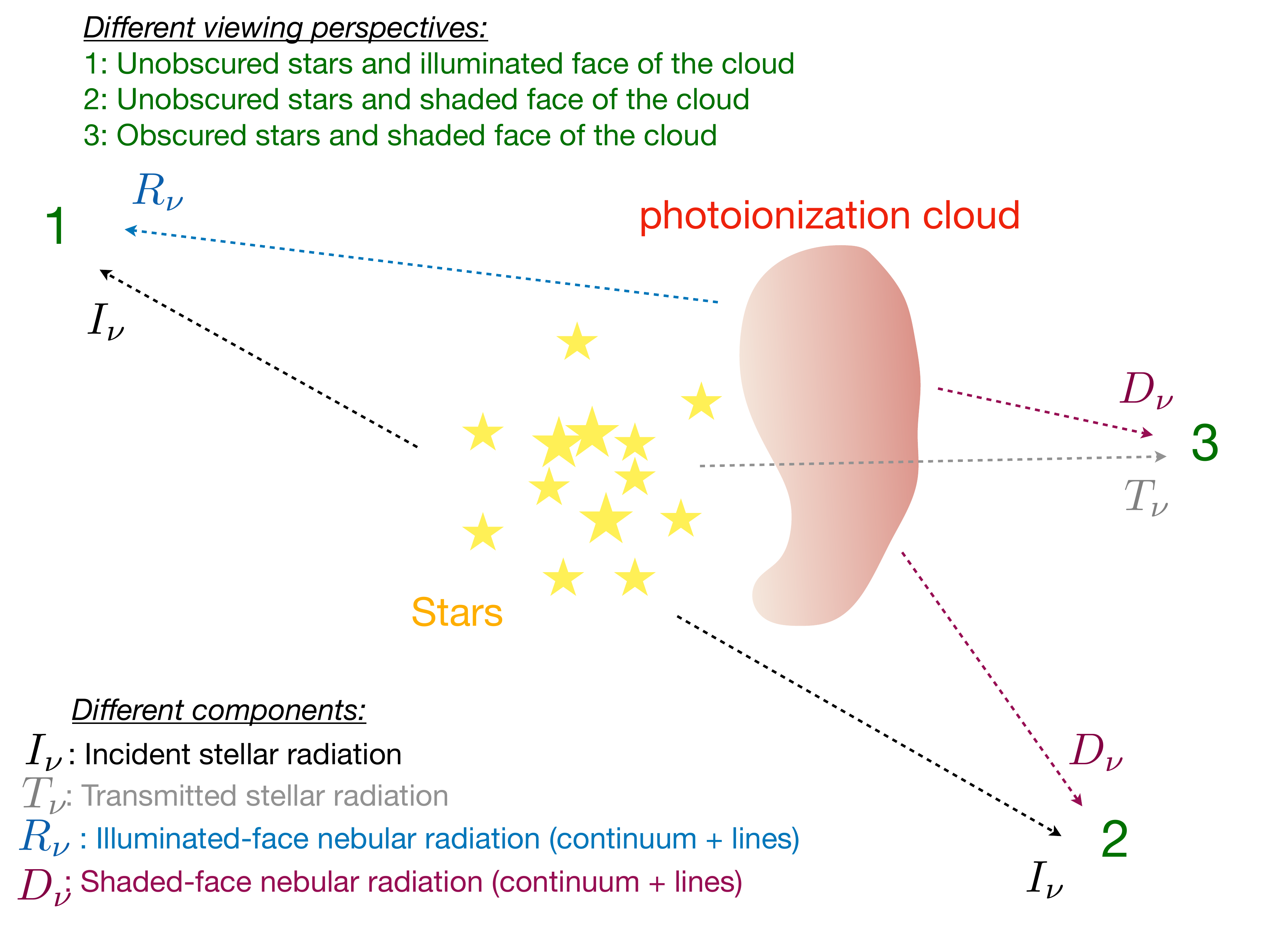}
    \caption{Effect of viewing angle on the observed spectrum. There are three generic viewing perspectives: (1) the observer sees unobscured stellar radiation $I_\nu$, as well as radiation $R_\nu$ emerging from the illuminated faces of photoionized clouds; (2) the observer sees unobscured stellar radiation $I_\nu$, as well as radiation $D_\nu$ emerging from the shaded faces of clouds; (3) the observer sees stellar radiation $T_\nu$ transmitted through the clouds, as well as radiation $D_\nu$ emerging from the shaded faces of the clouds.}
    \label{fig:specperspective}
\end{figure}

Observed continuum and line radiation depends on the observer's viewing angle. As illustrated in \reffig{specperspective}, there are three generic types of viewing angle --- the observer may see (1) light from exposed stellar photospheres and light emerged from the illuminated faces of the photoionized clouds; (2) light from exposed stellar photospheres and light emerged from the shaded faces of the clouds; (3) star light transmitted through the clouds and nebular light emerged from the shaded faces of the clouds. 

\texttt{Cloudy} calculations conveniently provide four different spectral components, whose superposition can well approximate a viewing-angle-averaged observed spectrum: direct star light $I_\nu$ (from BPASS templates), star light transmitted through clouds $T_\nu$, continuum and line nebular radiation emerging from illuminated cloud faces $R_\nu$, and that emerging from shaded cloud faces $D_\nu$. $T_\nu$ would exhibit spectral features due to absorption and scattering inside the clouds, while $R_\nu$ would include star light reflected backward by the clouds. Both $R_\nu$ and $D_\nu$ would include continuum radiation generated from the cloud interior (e.g. from radiative recombination). Optically thin emission lines would have equal fluxes in $R_\nu$ and $D_\nu$. For simplicity, we distinguish between direct light and transmitted light from the stars, and between illuminated and shaded faces of the clouds, but do not explicitly model the spectral dependence on a continuously varying viewing angle.

In the simplest model, we may consider a homogeneous population of clouds with fixed $\log P$ and $\log U$. The observed spectrum $F_\nu$ is given by
\begin{align}
    F_\nu = (1-z)\,I_\nu + z\,T_\nu + x \left( y\,R_\nu + (1-y)\,D_\nu \right),
\end{align}
where $0 \leqslant x \leqslant 1$ is the geometric fraction of star light processed by the clouds and $0 \leqslant z \leqslant 1$ is the fraction of stellar photospheres obscured by clouds when viewed along the observer's line of sight. We have also introduced an asymmetry parameter, the fraction of illuminated cloud face observed $0 \leqslant y \leqslant 1$, to quantify the probability of clouds located on the near side versus on the far side of the star cluster. If $y$ is closer to zero, clouds are preferentially located on the near side and we are more likely to see their shaded faces. If $y$ is closer to $1$, clouds are more likely located on the far side and we see their illuminated faces more. 

Emission line data favors a picture in which two distinct population of nebular clouds with significantly different ionization degrees and densities are present. One population of high ionization degree $\log U > -2$ and high pressure $\log P \approx 9$ are needed to explain the observed ${\rm [C III]}\lambda 1906/{\rm C III]}\lambda 1908$ and ${\rm [Si III]}\lambda 1883/{\rm Si III]}\lambda 1892$ line ratios. However, these clouds are too dense ($\log \nele \approx 5$) and too highly ionized to emit strong ${\rm [O II]}\lambda\lambda 3726,3729$, hinting at an additional population of lower $U$ and lower pressure.

HST continuum images in F555W, F606W and F814W show a two-component structure in Image 10 (see \reffig{astrometric} for an example). The extended component is $\sim 30-40\%$ as bright as the central component. Even in F390W and F410M, the extended component is similarly bright. The high relative brightness cannot be explained by recombination continuum. Reflected stellar UV light by dust grains only reaches $\sim 20\%$ even for a high dust albedo $a=0.65$ and a low forward scattering parameter $g=0.35$. Therefore, the extended continuum source must be predominantly stars on the outskirts of the cluster. 

\cite{Sharon2022SunburstLensModel} derived a source-plane upper bound on the cluster size $r_{\rm eff} \lesssim 20\,$pc. \cite{Vanzella2020IonizeIGM} and \cite{Vanzella2022SunburstEfficiency} reported marginal evidence from Image 10 seen in the rest-frame FUV continuum that the cluster has an effective angular radius $r_{\rm eff} = (0.06\pm 0.02)\arcsec$. Their results likely reflect the superposition of the central and extended stellar components.

The above considerations motivate us to consider a cluster model of two spatial components, one at the center and another extended. Either component has stars and is associated with a population of photo-ionized clouds. It is unclear whether the two stellar components are coeval, so our model allows two separate ages. One population of clouds closely surrounding the central component have an ionization parameter $U_1$ and a high pressure $P_1$ and process a fraction $x_1$ of the total star light with asymmetry parameter $y_1$. A second population of clouds, at larger distances from the cluster center, have a different ionization parameter $U_2$ and a lower pressure $P_2$, and process another fraction $x_2$ of the total star light with asymmetry parameter $y_2$. For this geometry, we assume that the low-pressure clouds are excited by ionizing radiation from both the central and the extended stellar components, while the high-pressure clouds are only excited by the central stellar component. Both populations are assumed to have the same metallicity, the same element abundance (except for nitrogen; see \refsec{abundances}), and the same grain content. 

The observed spectrum is given by the linear combination
\begin{align}
    F_\nu = & (1-z_1-z_2)\,I_\nu + z_1\,T_{1,\nu} + z_2\,T_{2,\nu} \nn
            & + x_1\,\left( y_1\,R_{1,\nu} + (1-y_1)\,D_{1, \nu} \right) \nn
            & + x_2\,\left( y_2\,R_{2,\nu} + (1-y_2)\,D_{2, \nu} \right) \nn
            & + (1-z_2)\,\hat I_\nu + z_2\,\hat T_{2,\nu} \nn
            & + x_2\,\left( y_2\,\hat R_{2,\nu} + (1-y_2)\,\hat D_{2, \nu} \right).
\end{align}
Contributions with and without a hat are powered by the central and extended stellar component, respectively. The subscript $_{1}$ and $_{2}$ indicates quantities that are associated with nebular reprocessing by the central and extended population of clouds, respectively. 

The geometric parameters are subject to the physical constraints $0 \leqslant x_1 + x_2 \leqslant 1 $ and $0 \leqslant z_1 + z_2 \leqslant 1 $. In this picture, a fraction $(1-z_1 - z_2)$ of the star light (including LyC radiation) reaches the observer along the line of sight without encountering any significant opacity, while the fraction $(1-x_1 - x_2)$ represents the {\it isotropic} escape fraction that star light escapes the cluster vicinity without being processed by optically thick gas.

\subsection{External Dust Reddening}
\label{sec:reddening}

Dust extinction due to Milky Way's ISM from UV through NIR is applied through the \texttt{dust\_extinction} package, using models from \cite{Fitzpatrick1999}. The model is defined by the parameter $R({\rm V}) = A({\rm V})/E({\rm B}-{\rm V})$, where we choose $R({\rm V})=3.1$, the mean value for MW extinction, and a normalization $E({\rm B}-{\rm V})=0.08$ according to the galactic dust maps measured by \cite{Schlegel1998}.

To account for possible dust reddening due to ISM in the host galaxy, we adopt the phenomenological reddening law of \cite{Reddy2015Mosdef}, which is empirically determined for $z\sim 1.4$--$2.6$ galaxies. This reddening law was used by \cite{Chisholm2019ExtragalacticMassiveStarPopulation} to analyze the rest-frame UV stellar spectral features of the LyC cluster. We normalize host ISM reddening using $\Ebv$, which we treat as a free parameter of the model.

\subsection{Element Abundances}
\label{sec:abundances}

From observed emission line strengths, we can measure gas phase abundance for four elements: (1) C/O from {\rm C III]}$\lambda\lambda$1906,1908; (2) N/O from {\rm N III]}$\lambda\lambda$1750,1752 or constrained by upper limits on {\rm [N II]}$\lambda\lambda$6548,6583; (3) Ne/O from {\rm [Ne III]}$\lambda\lambda$3869,3967; (4) Si/O from {\rm Si III]}$\lambda\lambda$1883,1892. As mentioned before, we set baseline abundance values for these elements in our \texttt{Cloudy} calculations (see \refsec{nebula}). Since for a typical abundance pattern oxygen forbidden lines are the main coolants in the photoionized gas, we assume that rescaling the abundance of a specific element, at the leading order, leads to the same rescaling of all lines associated with that element, while changes to cloud ionization and temperature can be neglected. In the fitting, we include one free rescaling factor for C, Ne and Si, respectively, which applies to both cloud populations. 

For nitrogen, the low-pressure clouds are located further away from the cluster than the high-pressure ones, so they are not necessarily equally enriched. We therefore introduce separate N abundance rescaling factors for both cloud populations. From X-shooter data we derive upper limits for the optical forbidden lines {\rm [N II]}$\lambda\lambda$6548,6583. These are suppressed if the photoionized gas has high density and high ionization degree. While low-density, low-ionization clouds contribute negligibly to {\rm N III]}$\lambda\lambda$1750,1752, their N enrichment can be constrained by non-detection of {\rm [N II]}$\lambda\lambda$6548,6583.

\subsection{Spectral Fitting}
\label{sec:spectralfitting}

Using model spectra, we compute observed fluxes for a list of HST filters (\reftab{photometry}), and a selected list of emission line ratios (\reftab{lineratio}). Assuming uncorrelated errorbars, the log likelihood used for fitting is minus one-half of the total chi square combining all photometric and emission line observables.

We include a total of 12 available HST filters, but exclude F390W, F410M, and F164N. Fluxes in F390W and F410M are sensitive to Ly$\alpha$ emission and damping wing attenuation, for which robust predictions are challenging due to the possibility of significant intervening HI columns along the line of sight. Neutral gas content physically far from the cluster may contribute to this column, about which data provide inadequate information to constrain the details. Other filters should be insensitive to these effects. F164N, a narrow band filter expected to be enhanced by H$\beta$ emission, may suffer from flux calibration issues, consistently underpredicting the flux expected from spectroscopic measurements. While our fitting does show a disagreement between the model predicted flux and data at the $2\sigma$ level, we found that inclusion of F164N resulted in only minor changes to the inferred physical parameters. Regardless, we choose to remove the filter from fitting to be conservative.

Fluxes in several filters (F126N, F160W and F167N) are expected to be significantly enhanced by one or several strong emission lines, such as ${\rm [O II]}\lambda\lambda3726,3729$, H$\beta$, and ${\rm [O III]}\lambda\lambda4959,5007$. We find that through strong line contributions HST photometry significantly constrains nebular properties independently of the emission line data.

H II gas glows in continuum light due to radiative recombination of hydrogen. This nebular continuum is proportional to the amount of LyC radiation processed by the photoionized cloud. Young starburst systems with $t_{\rm age}\lesssim 5\,$Myr have strong enough LyC radiation relative to the non-ionizing FUV radiation, and therefore the nebular continuum is a sizable contribution to the observed continuum flux \citep{Reines2010NebularContinuum, Byler2017NebularContinuum}. Recombination continuum has spectral discontinuities at the wavelengths of Lyman limit, Balmer limit, Paschen limit, and so on. The Balmer discontinuity at rest-frame $3645\,\AA$ is particularly relevant at the redshift of Sunburst because several HST filters probe opposite sides of the redshifted break and hence are sensitive to the size of this discontinuity, which provides information about $t_{\rm age}$, $Z$ and the electron temperature $\Te$. The medium filter F153M, redward of the break, happens to be unaffected by strong lines. We indeed detect flux deficit in that filter relative to neighboring filters.

\reftab{lineratio} lists 6 line ratios that we include in the analysis. To be conservative about element abundance uncertainties, we refrain from including line ratios that involve different metals except for measuring the abundance of certain metals. Compared to analyzing MUSE IFU data, we are relatively less confident about absolute flux calibration when analyzing X-shooter slit spectroscopy data. Hence, we refrain from using line ratios between MUSE detection and X-shooter detection. Line ratios based on X-shooter measurements always involve two lines with close wavelengths.

In particular, CIII]$\lambda\lambda$1906,1908 and SiIII]$\lambda\lambda$1883,1892 line ratios probe clouds of high ionization degree and are sensitive $\nele$ diagnostics in the high density regime $3 \lesssim \log \nele \lesssim 6$ \citep{Keenan1992CIIISiIII, Jaskot2016CIII}. The [OII]$\lambda\lambda$3726,3729 line ratio probes $\nele$ in clouds of low ionization degree, and is sensitive to a lower density range $1 \lesssim \log \nele \lesssim 4$ \citep{OsterbrockFerland1989text}. Ratios between the O II and O III lines, ${\rm [O II]} \lambda\lambda 3726,3729 / {\rm [O III]} \lambda\lambda 4959,5007$ \citep{KewleyDopita2002LineDiagnostics, KobulnickyKewley2004, Dors2011} in rest-frame optical and ${\rm [O II]} \lambda 2471 / {\rm [O III]}\lambda\lambda 1660,1666$ \citep{Kewley2019ARAAreview} in rest-frame UV, are sensitive to ionization degree and gas temperature, and simultaneously constrain the relative contributions to nebular radiation from the low-pressure and high-pressure clouds. The commonly used ratio {\rm [O III]}$\lambda\lambda$4959,5007 / {\rm H$\beta$} probes the gas metallicity and ionization parameter.

Since we combine observables of Image 9 and Image 10, uncertainty in the magnification ratio between the two must be taken into account. We treat the magnification ratio for both the central and the extended components as two free parameters with informative priors. This has been described in \refsec{muratio}. 

\begin{table*}
	\centering
	\caption{Nebular emission line ratios included in the spectral analysis.}
	\label{tab:lineratio}
	\begin{tabular}{ccl}
		\hline
		\hline
        Line ratio & Value \& uncertainty & Comment \\
        \hline
        {\rm Si III]}$\lambda$1892 / {\rm [Si III]}$\lambda$1883 & $2.42\pm0.97$& Probe density of high-pressure H II clouds \\
	{\rm C III]}$\lambda$1908 / {\rm [C III]}$\lambda$1906 & $2.32\pm0.23$& Probe density of high-pressure H II clouds  \\
        {\rm [O II]}$\lambda$2471 / {\rm [O III]}$\lambda\lambda$1660,1666 & $0.23\pm0.06$& Probe ratio between the high-pressure and low-pressure clouds \\
        {\rm [O II]}$\lambda$3729 / {\rm [O II]}$\lambda$3727 &$0.50\pm0.32$ & Probe density of low-pressure clouds  \\
        {\rm [O II]}$\lambda\lambda$3726,3729 / {\rm [O III]}$\lambda\lambda$4959,5007 & $0.047\pm0.012$& Probe ratio between the high-pressure and low-pressure clouds\\
        {\rm [O III]}$\lambda\lambda$4959,5007 / {\rm H$\beta$} & $12.38\pm4.40$& Ionization and metallicity diagnostics  \\
        \hline
        \hline
	\end{tabular}
\end{table*}

\section{Parameter Inference Results}
\label{sec:results}

\begin{figure*}[t]
    \centering
    \includegraphics[width=18cm,trim=0.25cm 0.25cm 0.25cm 0.25cm,clip]{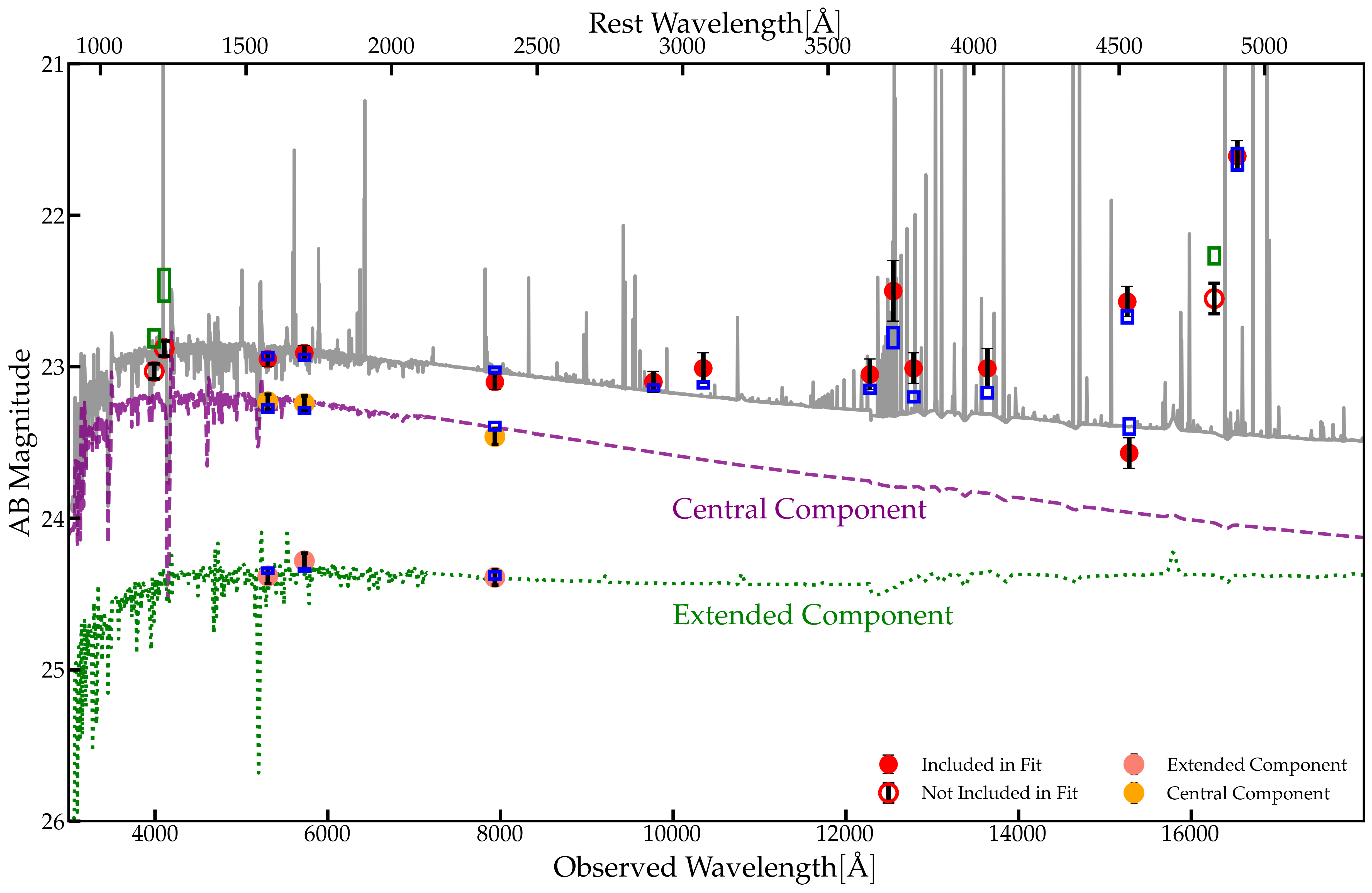}
    \caption{Best fit SED (grey) from \texttt{Cloudy} modeling. Co-plotted is the 68\% C.I. range of the model predicted filter magnitude (open boxes) and the observed magnitude (circles with error bars). Two of the bluest filters F390W and F410M (open circles and green boxes) are not used in the fit since the flux might have been significantly affected by Ly$\alpha$ line formation and damping wing scattering by an intervening H I column. The fitted models predict fluxes higher than observed in these two filters. The narrow-band F164N filter is also not used due to potential flux calibration issues. The observed magnitudes of the extended and central component from image 10 are shown demagnified by the median predicted magnification ratio for each component applied (pink and orange circles with errors bars), as well as 68\% C.I. range of the model predicted filter magnitudes demagnified in the same way (blue boxes). Best fit central (magenta) and extended (green) component continuum contributions are separately shown.}
    \label{fig:sed}
\end{figure*}

\begin{figure*}[t]
    \centering
    \includegraphics[width=18cm,trim=0.25cm 0.25cm 0.25cm 0.25cm,clip]{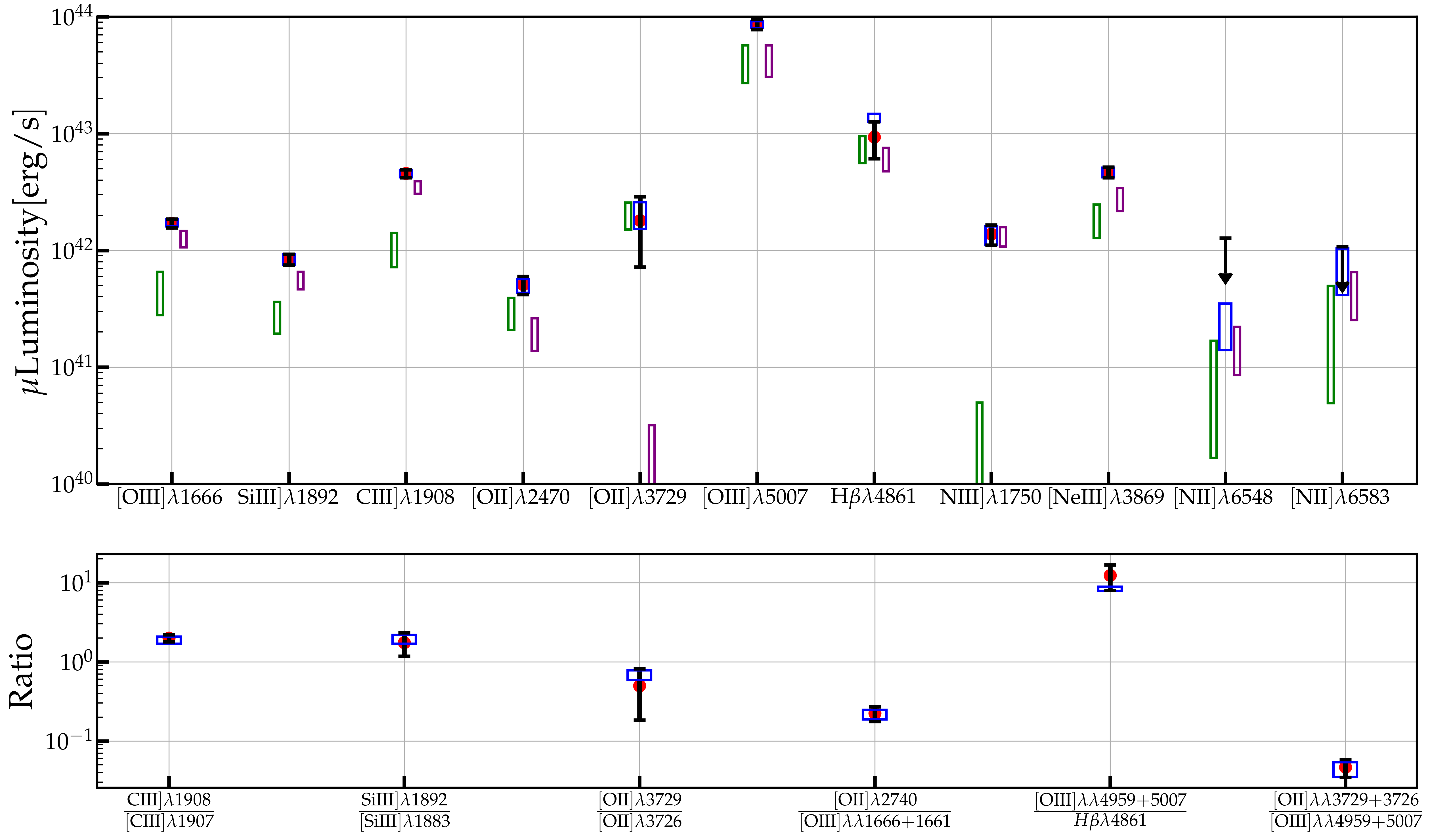}
     \caption{Top: 68\% C.I. for model predicted line fluxes (blue boxes) and observed line flux measured from MUSE/X-Shooter (circles with errorbars). The 68\% C.I. contributions from the high-pressure (purple boxes) and low-pressure (green boxes) clouds are shown separately. Bottom: 68\% C.I. for model predicted line ratios (blue rectangles) and observed line ratios measured from MUSE/X-Shooter (red markers).}
    \label{fig:lineflux}
\end{figure*}

We fit our \texttt{Cloudy} models to the observed photometry (\reffig{sed}), emission line fluxes and ratios (\reffig{lineflux}) using \texttt{PyMultinest} \citep{Feroz2009,Buchner2014}, with 1,000 live points running for 10,000 steps. The two-population model consists of 22 parameters in total: host dust reddening, metallicity, and the stellar age (for both stellar components); the pressure, ionization parameter and three geometric parameters for each population; the relative lensing magnification factor between the less magnified Image 9 (used for HST PSF photometry) and the more magnified Image 10 (used for emission line measurement) for both the central and extended components; the relative normalization of the two stellar components, overall normalization, and abundance rescalings for C, N (for both cloud populations), Ne and Si. The model is fit to 18 photometric measurements (Image 9 photometry for 12 HST filters, and Image 10 photometry for the central and extended components separately in 3 filters), 9 emission line fluxes, 6 emission line ratios, and 2 emission line non-detections; in sum 35 observables leading to 13 degrees of freedom overall. The best-fit model reproduces data with $\chi^2_\nu=1.85$.

\subsection{Star Cluster Properties}

\begin{figure}[t]
    \centering
    \includegraphics[width=\columnwidth]{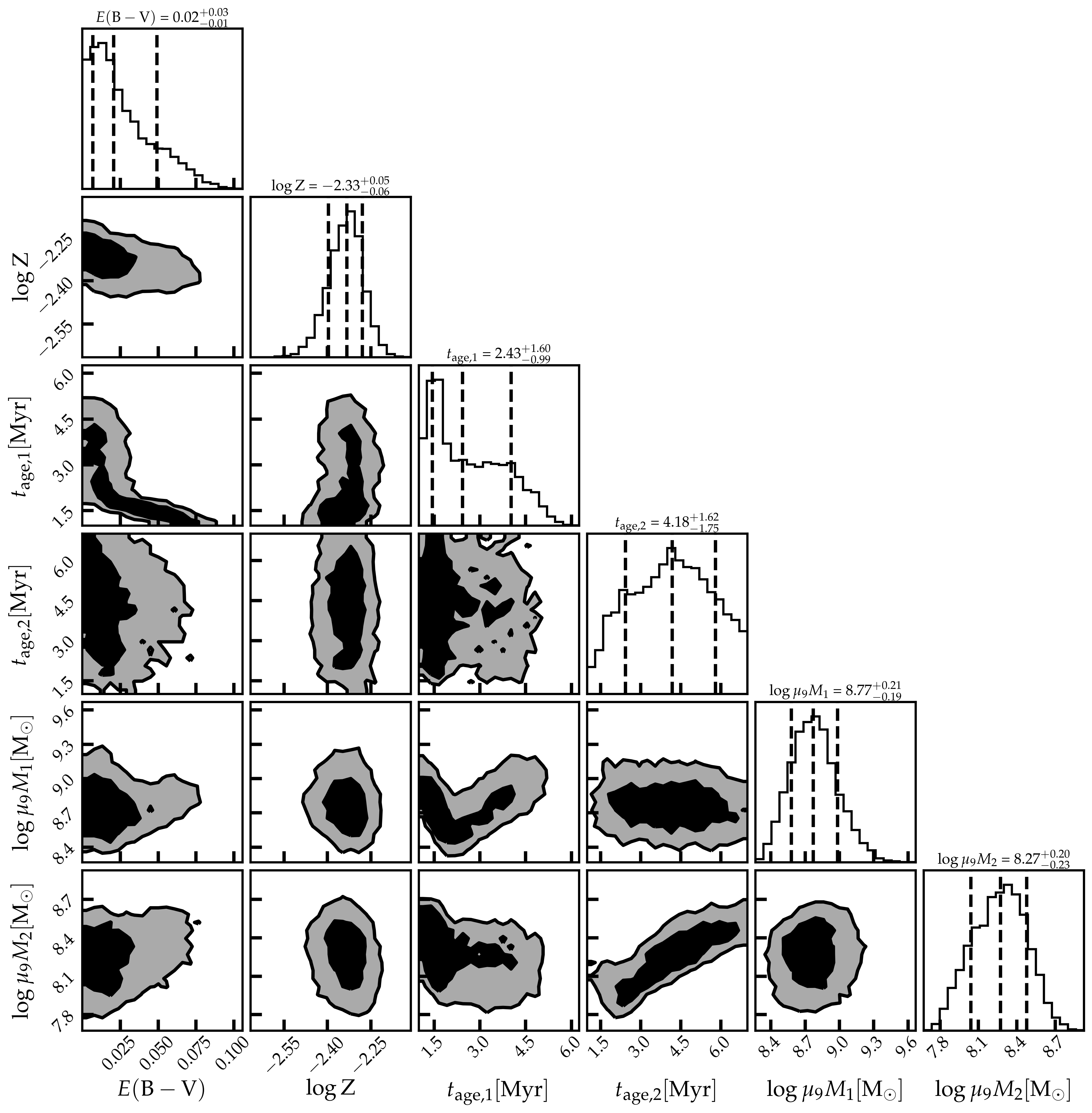}
    \caption{Posterior distributions for star cluster parameters. 1D histograms representing the posterior distributions are marked with the mean and 68\% confidence intervals (C.I.'s, dashed lines). 2D contours show the 50\% and 90\% levels for the density of points from \texttt{PyMultinest} fitting.
    Refer to the ``Two-Comp'' model in  \ref{tab:cloudtable} for the 68\% C.I.'s of the parameters.
    }
    \label{fig:starscorner}
\end{figure}

Properties of the star cluster are shown in \reffig{starscorner}. Assuming a single burst, we derive a cluster age $t_{\rm age}=2.4^{+1.6}_{-1.0}\,$Myr from the central component, which agrees with the independent UV spectroscopy analysis of \cite{Chisholm2019ExtragalacticMassiveStarPopulation} when the \texttt{BPASS} SED model is used, and is broadly consistent with the age requirement for very massive stars powering broad He II$\lambda$1640 wind emission as suggested by \cite{Mestric2023VMS}. 

The age constraint on the extended component is less tight, but the posterior peaks at $4\,$Myr. Data disfavors the extended component being older than $7$ Myr, and it is possible that the extended component is coeval with the central component. While our model suggests that the extended component may contribute as much as $\sim 25$\% of the ionizing flux, no extended source of escaping LyC radiation is seen in Image 10 in F275W \citep{Mainali2022OutflowLyCescape, Mestric2023VMS} (also see \reffig{astrometric}). The low density channels that allow LyC photons to pass freely may be connecting only to the central part of the cluster.   

We do not find any significant evidence for nonzero dust reddening caused by the host galaxy's ISM, with $\Ebv < 0.05$ at high confidence. The large-scale interstellar environment of the Sunburst galaxy toward the LyC cluster therefore does not appear dusty. This is in contrast to the result $\Ebv=0.15$ from \cite{Chisholm2019ExtragalacticMassiveStarPopulation} even though the same empirical reddening curve has been used. We are unable to explain the data with dust reddening as large as $\Ebv=0.15$ even when model details are tweaked. {While \cite{Chisholm2019ExtragalacticMassiveStarPopulation} usa an IMF of a similar shape, they use a lower 100$M_\odot$ upper mass cutoff than our analysis, however we found that adopting this upper mass cutoff did not significantly change our inferred reddening}. {There is a degeneracy between the nebular continuum, which renders the total FUV spectral slope redder (c.f. \reffig{sed}), and external dust reddening, but we find removal of the nebular continuum in our models results in only a modest increase in the inferred $\Ebv \sim 0.04$. \cite{Mainali2022OutflowLyCescape} measure reddening from stellar wind features in a similar approach to \cite{Chisholm2019ExtragalacticMassiveStarPopulation}, and find that the measured $\Ebv$ varies greatly between the different multiple images; the $\Ebv$ found for Image 9 and Image 10 are consistent with our results using those images. It is unclear which lensed images \cite{Chisholm2019ExtragalacticMassiveStarPopulation} used for their measurements, which are consistent with \cite{Mainali2022OutflowLyCescape}'s measurement of Image 5. \cite{Mainali2022OutflowLyCescape} also measure a dust reddening of $\Ebv\sim0.2$ via the balmer decrement, but use a different reddening law and fix the electron density and temperature to $n_e=10^2\,{\rm cm}^{-3}$ and $T_e=10^4\,\rm{K}$ respectively. However at our inferred densities of $ n_e \sim 10^{5}\,{\rm cm}^{-3}$, the balmer decrement is sensitive to density and temperature \citep{OsterbrockFerland1989text}, and the unreddened balmer decrement predicted by our models is much higher than those assumed in \cite{Mainali2022OutflowLyCescape}, resulting in a lower $\Ebv$.}

{We infer a gas-phase metallicity of $Z\sim 0.2\,Z_{\odot}$, significantly lower than the stellar metallicites of  $\sim 0.5-0.6\,Z_{\odot}$ and  $\sim 0.4\,Z_{\odot}$ inferred by \cite{Chisholm2019ExtragalacticMassiveStarPopulation} and by \cite{Mainali2022OutflowLyCescape} respectively via the CIV$\lambda 1550$ and NV$\lambda 1240$ P Cygni stellar wind features. This is somewhat in contrast to other cosmic noon star forming galaxies, which usually have similar or slightly lower stellar metallicites when compared to the gas-phase \citep{Sommariva2012StellarMetallicity}. \cite{Mestric2023VMS} infer a gas-phase metallicity of $Z \sim 0.4\,Z_{\odot}$ via the N2 index \citep{Marino2013N2Index}, but the lack of a clear $[\rm{N II}] \lambda6583$ detection means this may instead be an upper limit.}

For the IMF we choose (c.f. \refsec{stars}), we measure a stellar mass $\log(\mu_9\,M_{\star, 1}/M_\odot)=8.8^{+0.2}_{-0.2}$ for the central component, and $\log(\mu_9\,M_{\star, 2}/M_\odot)=8.3^{+0.2}_{-0.2}$ for the subdominant extended component. The precise mass is unknown due to the large uncertainty in the magnification factor of Image 9, $\mu_9$, as multiple independent lens modeling efforts \citep{Vanzella2020IonizeIGM, Pignataro2021SunburstLensModel, Diego2022godzilla, Sharon2022SunburstLensModel} have obtained magnification factors of the various lensed images of the cluster that are different from each other and are discrepant with the observed flux ratios. A plausible range for the magnification of Image 9 might be $\mu_9 = 10$--$30$. Just the central component has $\sim 5 \times 10^6\,M_\odot$ of stars even if Image 9 has $\mu_9=100$, which is one or two orders of magnitude more massive than any exposed young super star clusters seen in the local Universe. This is in agreement with the inference of \citep{Vanzella2020IonizeIGM, Vanzella2022SunburstEfficiency} and supports the idea that such a system will likely evolve to a massive globular cluster. 

We repeated our analysis with a top-heavy IMF with $\xi(m)\propto m^{-2}$ for $m>0.5\,M_\odot$. No significant changes to the model parameters were seen, except that the inferred cluster mass decreased by a factor of $\sim 2$ (see \reftab{cloudtable}). 
This is because at $t_{\rm age} \lesssim 4\,$Myr the cluster ionizing output is completely dominated by the most massive stars. In particular, the central component of the young super star cluster has a huge output of LyC photons, $\log(\mu_9\,Q({\rm H^0_{1}}))=55.5^{+0.3}_{-0.2}$. Barring the uncertainty in $\mu_9$, this result is insensitive to the IMF slope. {We also decreased the upper mass cutoff to $100\,M_\sun$, which resulted in a more sharply peaked age posterior at $1.5$--$2\,$Myr, but we did not find any significant differences in our inferred parameters}.

\subsection{Nebula Properties}

\begin{figure}[t]
    \centering
    \includegraphics[width=\columnwidth]{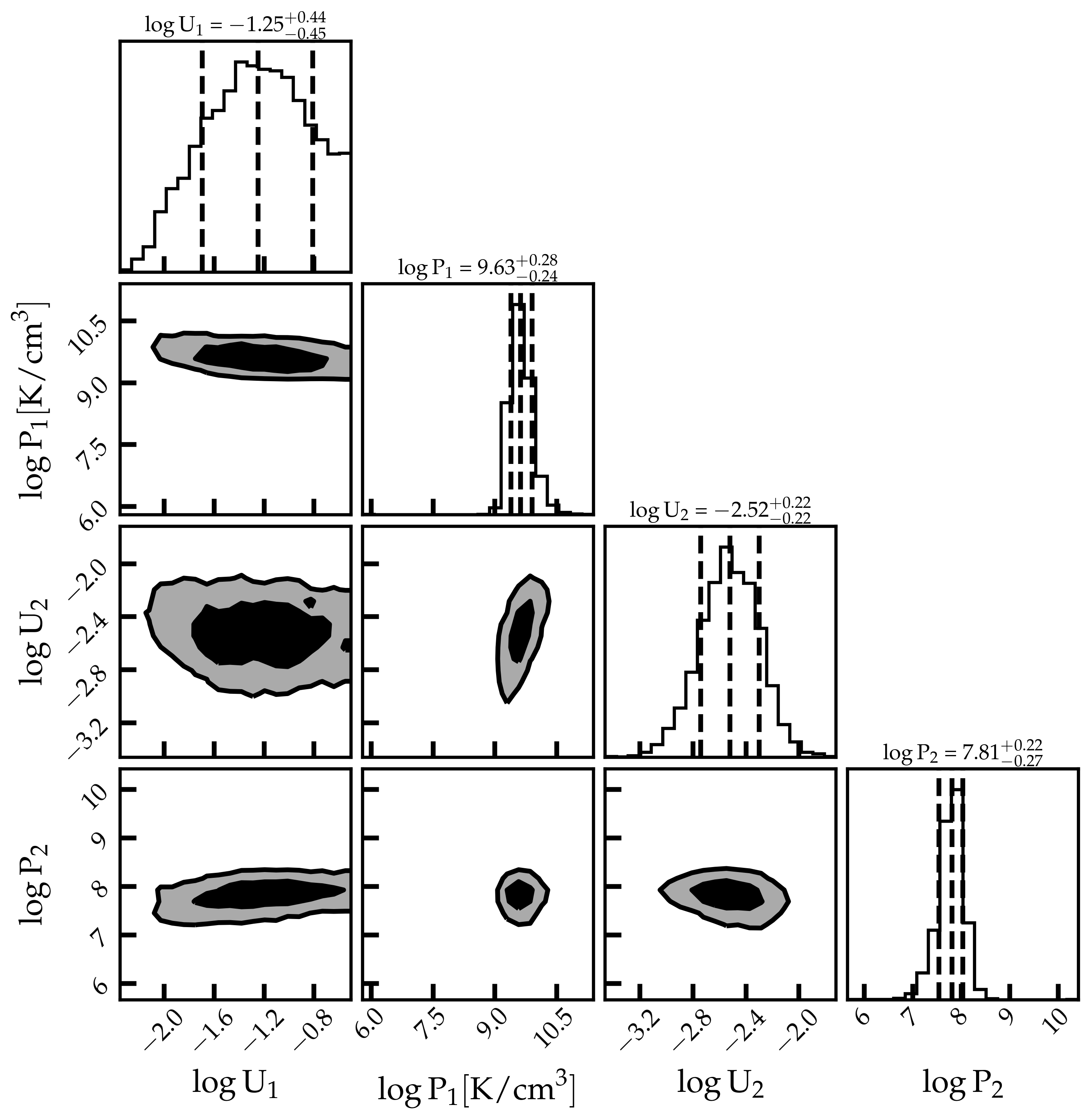}
    \caption{Posterior distributions for the measured physical parameters of the two cloud populations. Contours levels and histogram confidence intervals follow Fig.~\ref{fig:starscorner}. Parameter 68\% C.I.'s are tabulated in \reftab{cloudtable}.}
    \label{fig:cloudcorner}
\end{figure}

\reffig{cloudcorner} shows inferred parameters describing the photoionized clouds: cloud pressure $P$, ionization parameter $U$, and the ionizing flux incident onto the cloud $\Phi({\rm H}^0)$. These intensive nebula parameters can be inferred independently of the uncertain lensing magnification factor.

The model has two cloud populations with well-constrained but distinct pressure values. The high-pressure clouds, which is the primary source for emission lines from high ionization species, have $\log P_1 = 9.6^{+0.3}_{-0.2}$ (pressure in units of ${\rm K}\,{\rm cm}^{-3}$). Their ionizing parameter is high $\log U_1 =-1.25^{+0.4}_{-0.4}$. The H II gas pressure is constrained to be high because {\rm C III]}$\lambda\lambda$1906,1908 line ratio indicates $\log \nele \approx 5$.
The intensity of ionizing photons striking the cloud surface (in units of ${\rm cm}^{-2}\,{\rm s}^{-1}$) has a high value $\log \Phi_1({\rm H}^0) = 14.3^{+0.3}_{-0.3}$. 

A second population of low-pressure clouds account for the strong [O II]$\lambda\lambda$3726,3729 doublet. They have a lower pressure $\log P_2 = 7.8^{+0.2}_{-0.3}$, a lower ionization parameter $\log U_2 = -2.5^{+0.2}_{-0.2}$, and are irradiated by a weaker ionizing flux $\log \Phi_2({\rm H}^0) = 11.5^{+0.3}_{-0.3}$.

Line profile analysis performed in \cite{Mainali2022OutflowLyCescape} revealed that {\rm [O III]}$\lambda$5007 emission from the LyC cluster consists of two components with distinct kinematics: a narrow component with FWHM=$100\,{\rm km}\,{\rm s}^{-1}$ and a broader component with FWHM=$300\,{\rm km}\,{\rm s}^{-1}$. Since our model predicts that strong UV metal lines such as {\rm C III]}$\lambda$1908 are predominantly from the high-pressure nebula, and that those lines appear consistent with FWHM=$100\,{\rm km}\,{\rm s}^{-1}$ rather than FWHM=$300\,{\rm km}\,{\rm s}^{-1}$, we surmise that the narrow and broad line components correspond to the high- and low-pressure nebulae, respectively.

\subsection{Geometric Parameters}

\begin{figure}[t]
    \centering
    \includegraphics[width=\columnwidth]{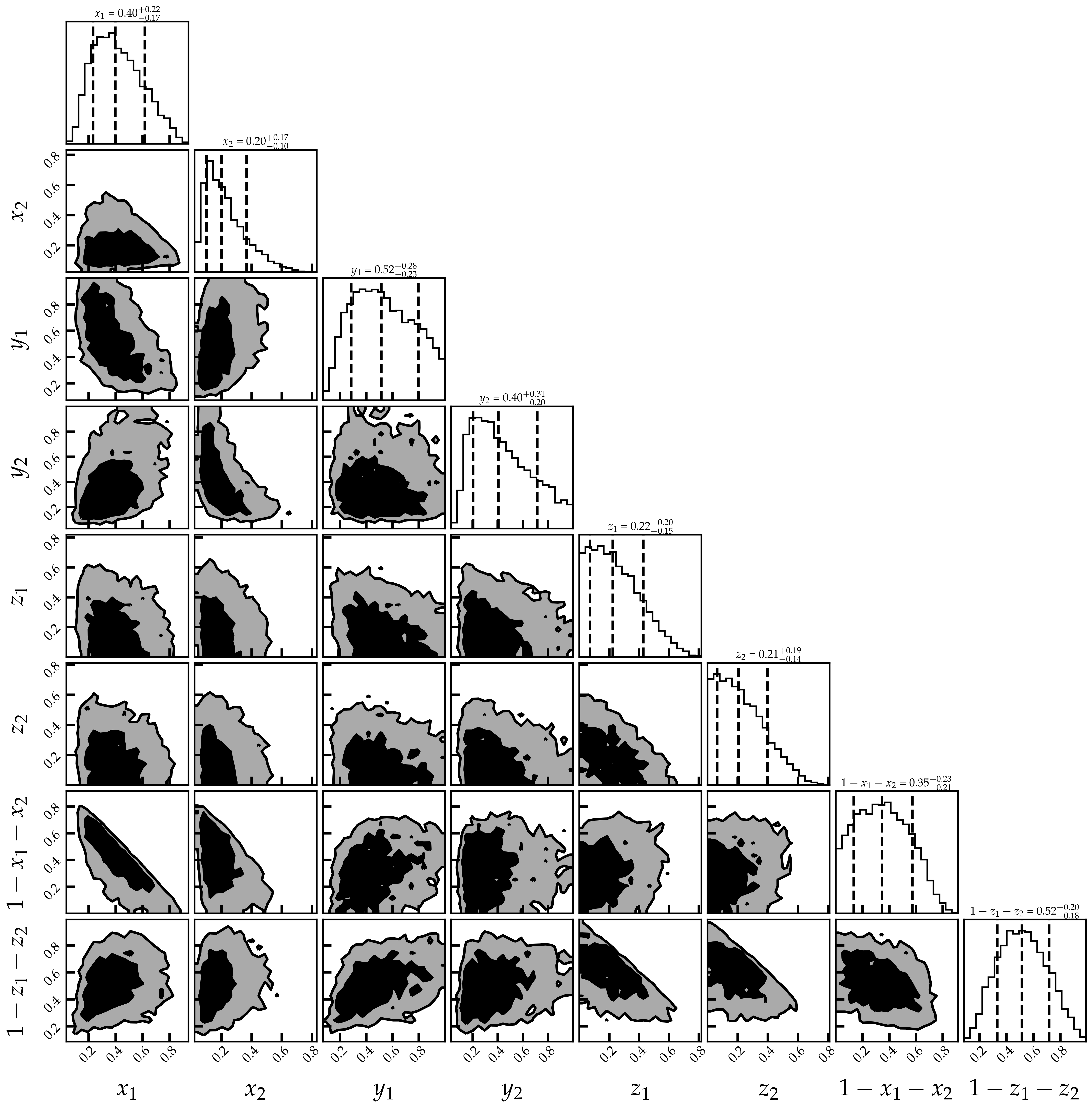}
    \caption{Posterior distributions for the measured geometric parameters of the two clouds, including covering factors, obscuring factors, and viewing angles. Contours levels and histogram confidence intervals follow Fig.~\ref{fig:starscorner}. 
    Refer to the ``Two-Comp'' model in \reftab{cloudtable} for the 68\% C.I.'s of the parameters.
    }
    \label{fig:geocorner}
\end{figure}

Our analysis simultaneously constrain the spatial geometry of the surrounding photoionized clouds. Information about these mainly come from the relative fluxes between the observed lines and continuum, the spectral difference between the illuminated and shaded cloud faces, as well as the amount of reddening for stellar light transmitted through any obscuring clouds that intervene the light of sight compared to the incident stellar light. 

The results for geometric parameters are shown in \reffig{geocorner}. About $x_1 = (40^{+22}_{-17})\%$ of the cluster's ionizing output is processed by high-pressure clouds where high ionization emission lines are powered, and a remaining $x_2 = (20^{+17}_{-15})$\% is processed by low-pressure clouds. While the geometry parameters have broad posterior distributions, our fitting results imply that a substantial fraction $(1-x_1-x_2) = (35^{+23}_{-21})$\% of the ionizing output leaves the cluster vicinity in various directions without being processed by opaque clouds. 

On the other hand, along the specific line of sight toward the observer, an even higher fraction $(1-z_1-z_2)=(52^{+20}_{-18})$\% of the ionizing radiation from the stars escapes the system. This picture is fully consistent with previous study carried out by \cite{RiveraThorsen2019Sci}, who suggest a line-of-sight LyC escape fraction $40$--$90$\% if dust absorption within the host galaxy is negligible.

Since we assume that the photoionized clouds have a high H column density and have dust grains, illuminated cloud faces and shaded cloud faces are distinguishable because continuum and line radiation emerging from the shaded cloud faces is severely suppressed, especially at rest-frame UV wavelengths. In our results, both $y_1$ and $y_2$ have a broad posterior consistent with $y_1 = 1/2$ and $y_2 = 1/2$. This is compatible with the picture in which the photoionized clouds, of both populations, are isotropically distributed around the star cluster. In our opinion, this is a simple, natural scenario. If data preferred a situation where clouds were predominantly seen from the illuminated faces or from the shaded faces, we would have to either attribute that to chance, or would have to invoke additional physical explanations for such geometric asymmetry. There is insufficient information to determine whether the photoionized clouds are shattered into large quantities of cloudlets, in which case the line-of-sight escape fraction $(1-z_1-z_2)$ may equal the isotropic escape fraction.

\subsection{Element Abundances}

\begin{figure}[t]
    \centering
    \includegraphics[width=8.25cm,trim=0.25cm 0.25cm 0.25cm 0.25cm,clip]{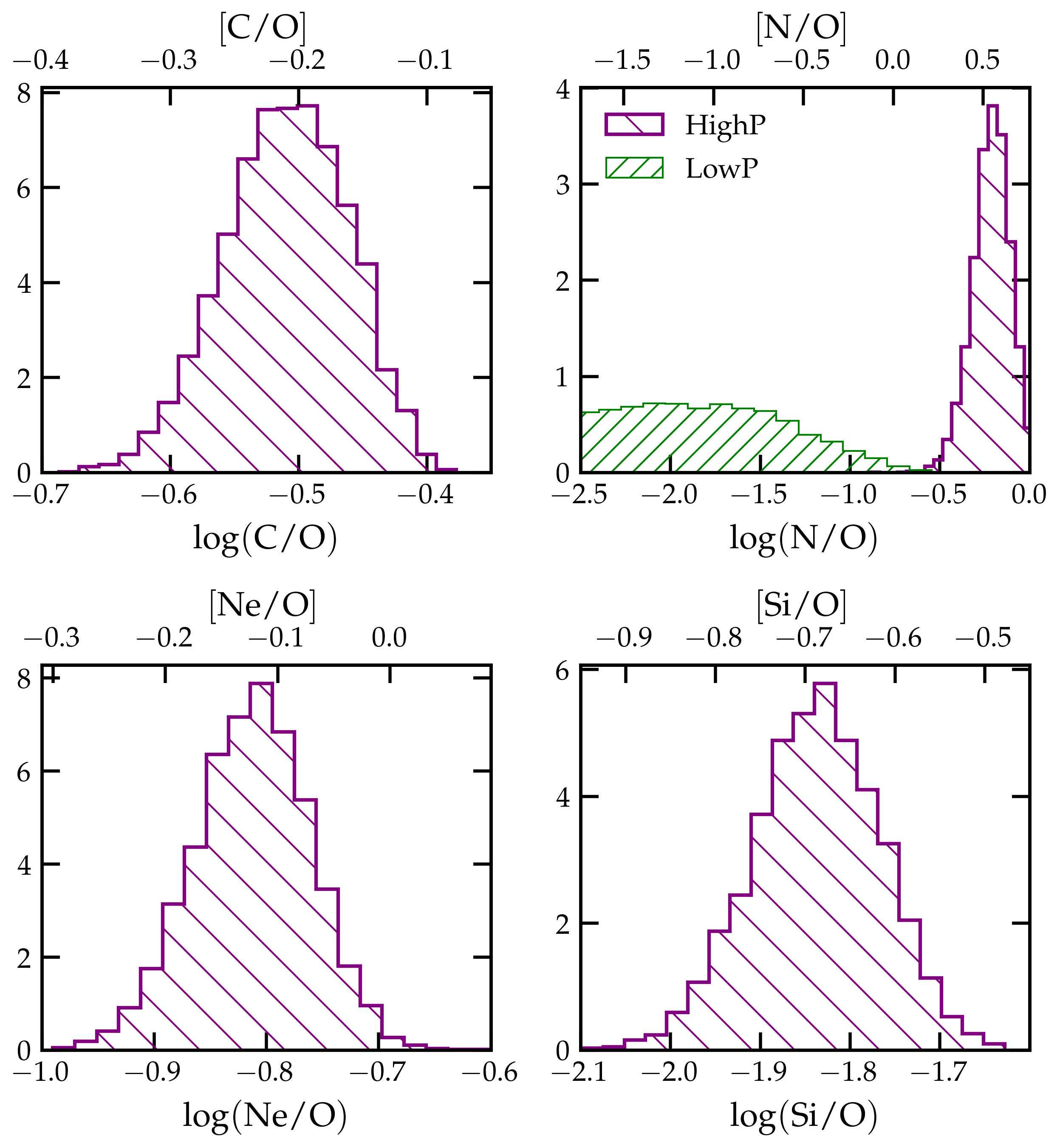}
    \caption{Posterior distributions for the gas-phase abundances of C, N, Ne and Si (arbitrary units for the vertical axis). {We plot the absolute abundance on the bottom axis, and abundance relative to solar on the top axis}. In the upper right panel, we observe high elevation of N abundance in the high-pressure clouds (purple) compared to the median ISM abundance at our inferred metallicity $Z \approx 0.2$--$0.3\,Z_\odot$, which indicates enrichment by massive star ejecta. By contrast, the low-pressure, more distant clouds (green) cannot be as much enriched and are consistent with uncontaminated. We also observe significant Si depletion which is evidence for grain formation. C and Ne appear to have normal abundances. {The results are also shown in \reftab{abundances}.}}
    \label{fig:abundances}
\end{figure}

Under the assumption that $[{\rm O/H}] = \log(Z/Z_\odot)$ for a stellar metallicity $Z$ and with our fiducial abundance pattern, we measure $Z/Z_\odot = 0.22\pm 0.03$. This is lower than the results of \cite{Chisholm2019ExtragalacticMassiveStarPopulation}, $Z/Z_\odot = 0.5$--$0.6$, which was inferred from rest-frame UV spectral features of stellar winds.
We consistently infer a gas metallicity $Z\approx (0.2-0.3)\,Z_\odot$ when modeling details are varied (c.f. \reftab{cloudtable}).

The picture of dusty photoionized clouds is corroborated by the inferred gas-phase Si depletion by a factor $\sim 4$--$5$ (\reffig{abundances}). If seeds are present in the first place, grain growth should be efficient at high densities $\log\nele =4$--$5$ and a cool gas temperature at a few thousand K, even under strong external UV irradiation. For example, dust grains are observed to rapidly form in colliding wind shells of the Wolf-Rayet binary system WR140 \citep{Han2022NatureWR140DustShell, Lau2022NatAsWR140DustShellsJWST}.

Mg II$\lambda\lambda$ 2795,2803 doublet and [Fe IV]$\lambda\lambda$2829,2835 doublet are transitions from ionic ground levels. Photoionization calculation shows that strong emissions should emerge from the illuminated cloud face. Inspecting the the X-shooter data, we nevertheless detect only weak Mg II$\lambda\lambda$ 2795,2803 emission, and find no sign for [Fe IV]$\lambda\lambda$2829,2835 emission. The plausible explanation is that gas-phase Si, Mg and Fe are all depleted due to grain formation. This further supports the model of dusty photoionized clouds.

We have detected strikingly elevated nitrogen abundance $\log({\rm N/O})=-0.21^{+0.10}_{-0.11}$ for the high-pressure clouds, a factor of $\sim 12$--$19$ higher than the typical ISM abundance $\log({\rm N/O}) \simeq -1.4$ at $Z/Z_\odot \simeq 0.2$ \citep{Molla2007NitrogenEvolution}. By contrast, no evidence of enrichment is seen for the low-pressure clouds, which have $\log({\rm N/O})\lesssim-1.31$. This is evidence for localized self-enrichment by massive star ejecta (discussed in \refsec{nitrogen} with more details).

No dramatic abundance anomalies are noticeable for carbon $\log({\rm C/O})=-0.51^{+0.05}_{-0.05}$ and for neon $\log({\rm Ne/O})=-0.81^{+0.05}_{-0.05}$. The unremarkable Ne/O ratio disfavors significant gas-phase O depletion. Abundance results are summarized in \reffig{abundances} and \reftab{abundances}.

\begin{table}
	\centering
	\caption{Inferred element abundances of carbon, nitrogen, neon and silicon. Also refer to the ``Two-Comp'' model in \reftab{cloudtable}.}
	\label{tab:abundances}
	\begin{tabular}{rccl}
		\hline
		\hline
		Abundance & Value & Solar & Important lines \\
		\hline
        $\log({\rm C/O})$ & $-0.51^{+0.05}_{-0.05}$ & $-0.30$ & {\rm C III]}$\lambda\lambda$1906,1908 \\
        \footnote{Nitrogen abundance for high-pressure clouds.}
        $\log({\rm N/O})$ & $-0.21^{+0.10}_{-0.11}$ & $-0.76$ & {\rm N III]}$\lambda\lambda$1750,1752 \\
        \footnote{Nitrogen abundance for low-pressure clouds.}
        $\log({\rm N/O})$ & $\lesssim-1.31$ & $-0.76$ & {\rm [N II]}$\lambda\lambda$6548,6583 \\
        $\log({\rm Ne/O})$ & $-0.81^{+0.05}_{-0.05}$ & $-0.69$ & {\rm [Ne III]}$\lambda\lambda$3869,3967 \\
        $\log({\rm Si/O})$ & $-1.84^{+0.07}_{-0.07}$ & $-1.15$ & {\rm Si III]}$\lambda\lambda$1883,1892 \\
		\hline
        \hline
	\end{tabular}
\end{table}

\subsection{Altering Model Assumptions}

We study how robust our parameter inference is by modifying some of the basic modeling assumptions.  
We study three alternate model assumptions. Parameter inference results for these modified models, together with our default model (``Two-Comp.'') already explained in previous sections, are tabulated in \reftab{cloudtable}.

For the first alternative model (``Top-heavy''), we consider a top-heavy IMF for which high mass stars are up-weighted by having $\xi(m) \propto m^{-2}$ for $m > 0.5\,M_\odot$, but retains the same IMF slope for lower mass stars, the same mass cutoff, and instant star formation. This model has a comparable reduced chi square. Compared to the default model, the inferred parameters are consistent within $1\sigma$, with only exception being the magnified cluster stellar mass for both the central and extended components, which are a factor of two smaller. This is easily understandable because high mass stars dominate all light observables at $\lesssim 4\,$Myr; the number of low mass stars, which dominate the stellar mass, is necessarily estimated by extrapolation following the assumed IMF. The same mass-IMF degeneracy is expected for other forms of top-heavy IMF.

Even though an instant starburst can be a good approximation for a localized starburst, we explore the alternative scenario of continuous star formation at a constant rate as the second alternative model (``Cont. SF''). This model has a reduced chi square comparable to that of the default model, and nearly all inferred parameters change by less than $1\sigma$. The notable exception is the central cluster age, which favors $4.3\,$Myr but with larger uncertainties ($\pm 1-2\,$Myr). Similarly, the extended cluster age favors $4.7\,$Myr with large uncertainties. This suggests observable features sensitive to $t_{\rm age}$ in the burst model are no longer constraining for continuous star formation. While our fitting does not show a significant preference for either an instant starburst or continuous star formation, \cite{Chisholm2019ExtragalacticMassiveStarPopulation} found that constant star formation did not reproduce the observed {\rm C IV}$\lambda$1550 and N V$\lambda$1240 P-Cygni features of massive star winds as well as instant starburst does.

Finally, we also test a simplified model without photometric constraints from Image 10, and without a second stellar population (``One-Comp.''). This model retains only a single free parameter for the magnification ratio between Images 9 and 10, which is applied uniformly across both the high-pressure and low-pressure cloud emission in order to fit to emission line fluxes measured from Image 10. With the removal of 6 observational constraints from Image 10 photometry and 3 free parameters corresponding to the extended stellar component, this model instead has only 11 degrees of freedom.  We find this model provides a similarly good $\chi^2_\nu$ to the two-stellar component model, and the inferred physical parameters are in agreement with the two-stellar component model. The inferred cluster age for the single stellar component, $t_{\rm age} = 3.5^{+0.9}_{-0.9} {\rm Myr}$, is broadly in agreement with the central component from the two-stellar component model. This single stellar component has inferred magnified mass $\log(\mu_9\,M_\star) = 8.8^{+0.2}_{-0.2}$, and magnified ionizing photon rate $\log(\mu_9 Q({\rm H}^{0}))=55.4^{+0.2}_{-0.2}$, also in agreement with the central component of the two-stellar component model. The high-pressure clouds see small variations in their ionization parameter and pressure, with $\log U_1 = -1.2^{+0.3}_{-0.4}$ and $\log P_1 = 9.8^{+0.7}_{-0.4}$, however these are well within $1\sigma$ of the two-stellar component model. The uniform magnification ratio inferred is $\mu_{10}/\mu_9 = 3.40^{+0.34}_{-0.34}$, which is consistent at the $2\sigma$ level with the total magnification ratio measured photometrically in the F555W, F606W, and F814W filters as described in \refsec{muratio}. The physical parameters for the low-pressure clouds, the geometric factors, and elemental abundances all see negligible changes between the two models. 

We conclude that our most important qualitative findings about the cluster (e.g. age, metallicity, external dust reddening) and about the dense photoionized clouds (e.g. high pressure and high ionization, nitrogen enhancement, silicon depletion) are robust against tweaking of detailed model assumptions.


\begin{table*}
	\centering
	\caption{Physical parameters derived for the two-stellar population (``Two-Comp.''), top-heavy IMF (``Top-heavy''), single-stellar population (``One-Comp.''), and continous star-formation (``Cont. SF'') models. The ``Two-Comp.'' model is the default model of this work. Pressures are in units of ${\rm K}\,{\rm cm}^{-3}$, photon fluxes in units of ${\rm cm}^{-2}\,{\rm s}^{-1}$, and radii in units of ${\rm cm}$. $\mu_9$ is the magnification factor of Image 9.}
	\label{tab:cloudtable}
	\begin{tabular}{c|l|cccc}
		\hline
            \hline
		 Parameter & Meaning & Two-Comp. & Top-heavy & One-Comp.$^a$ & Cont. SF \\
		\hline
		$\chi^{2}_{\nu}$ & Reduced chi square & 1.85 & 1.82 & 1.89 & 1.96\\
		$\chi^{2}$ & Chi square & 24.08 & 23.68 & 18.92 & 25.51\\
            \hline
		$\Ebv$ & Host ISM dust reddening & $0.02^{+0.03}_{-0.01}$ & $0.02^{+0.02}_{-0.01}$ & $0.01^{+0.01}_{-0.01}$ & $0.04^{+0.02}_{-0.02}$ \\
		$\log Z^{b}$ & Absolute metallicity & $-2.33^{+0.05}_{-0.06}$ & $-2.31^{+0.05}_{-0.06}$ & $-2.20^{+0.06}_{-0.06}$ & $-2.31^{+0.06}_{-0.07}$ \\
		$t_{\rm age,1}$ & Cluster age of central stars [Myr] & $2.43^{+1.60}_{-0.99}$ & $3.33^{+1.24}_{-1.77}$ & $3.47^{+0.92}_{-0.97}$ & $4.33^{+1.68}_{-1.85}$ \\
		$t_{\rm age,2}$ & Cluster age of outskirt stars [Myr] & $4.18^{+1.62}_{-1.75}$ & $4.18^{+1.70}_{-1.69}$ & --- & $4.73^{+1.48}_{-1.70}$ \\
        \hline
		$\log U_1$ & Ionization parameter of high-pressure clouds & $-1.25^{+0.44}_{-0.45}$ & $-1.35^{+0.48}_{-0.46}$ & $-1.19^{+0.47}_{-0.51}$  & $-1.24^{+0.45}_{-0.53}$ \\
        $\log U_2$ & Ionization parameter of low-pressure clouds & $-2.52^{+0.22}_{-0.21}$ & $-2.50^{+0.22}_{-0.22}$ & $-2.41^{+0.21}_{-0.28}$  & $-2.51^{+0.23}_{-0.22}$ \\
		$\log P_1$ & Total pressure of high-pressure clouds [${\rm K}\,{\rm cm}^{-3}$] & $9.63^{+0.28}_{-0.24}$ & $9.62^{+0.27}_{-0.25}$ & $9.79^{+0.71}_{-0.40}$  & $9.69^{+0.34}_{-0.29}$ \\
		$\log P_2$ & Total pressure of low-pressure clouds [${\rm K}\,{\rm cm}^{-3}$] & $7.81^{+0.22}_{-0.27}$ & $7.78^{+0.25}_{-0.29}$ & $7.86^{+0.23}_{-0.27}$  & $7.84^{+0.23}_{-0.27}$ \\
            \hline
        $\log(\mu_9\,M_{\star,1})$ & (Magnified) stellar mass of central stars [${\rm M_\odot}$] & $8.77^{+0.21}_{-0.19}$ & $8.41^{+0.25}_{-0.26}$ & $8.85^{+0.20}_{-0.21}$  & --- \\
         $\log(\mu_9\,M_{\star,2})$ & (Magnified) stellar mass of extended stars [${\rm M_\odot}$] & $8.27^{+0.20}_{-0.19}$ & $7.94^{+0.26}_{-0.34}$ & ---  & --- \\
        $\log(\mu_9\,{\rm SFR}_{1})$ & (Magnified) SFR of central stars [${\rm M_\odot}/\rm{yr}$] & --- & --- & ---  & $2.13^{+0.30}_{-0.19}$ \\
        $\log(\mu_9\,{\rm SFR}_{2})$ & (Magnified) SFR of extended stars [${\rm M_\odot}/\rm{yr}$] & --- & --- & ---  & $1.49^{+0.19}_{-0.13}$ \\
		$\log(\mu_9 Q({\rm H}_{1}^{0}))$ & (Magnified) ionizing output of central stars [${\rm s}^{-1}$] & $55.46^{+0.28}_{-0.23}$ & $55.39^{+0.29}_{-0.21}$ & $55.38^{+0.18}_{-0.15}$  & $55.59^{+0.21}_{-0.16}$ \\
  		$\log(\mu_9 Q({\rm H}_{2}^{0}))$ & (Magnified) ionizing output of outskirt stars [${\rm s}^{-1}$] & $54.66^{+0.22}_{-0.23}$ & $54.67^{+0.19}_{-0.21}$ & ---  & $54.98^{+0.15}_{-0.13}$ \\
		$\log\Phi_1({\rm H}^{0})$ & Ionizing flux incident on high-pressure clouds [${\rm s}^{-1}\,{\rm cm}^{-2}$] & $14.34^{+0.28}_{-0.26}$ & $14.29^{+0.28}_{-0.28}$ & $14.50^{+0.68}_{-0.39}$  & $14.41^{+0.32}_{-0.30}$ \\
		$\log\Phi_2({\rm H}^{0})$ & Ionizing flux incident on low-pressure clouds [${\rm s}^{-1}\,{\rm cm}^{-2}$] & $11.54^{+0.28}_{-0.30}$ & $11.53^{+0.31}_{-0.31}$ & $11.69^{+0.32}_{-0.39}$  & $11.58^{+0.31}_{-0.30}$ \\
		$\log (R_1\mu_9^{1/2})$ & Distance of high-pressure nebula [{\rm cm}] multiplying $\sqrt{\mu_9}$ & $20.02^{+0.17}_{-0.17}$ &  $20.02^{+0.18}_{-0.17}$ & $19.89^{+0.21}_{-0.35}$  & $20.06^{+0.19}_{-0.19}$ \\
		$\log (R_2\mu_9^{1/2})$ & Distance of low-pressure nebula [{\rm cm}] multiplying $\sqrt{\mu_9}$ & $21.43^{+0.20}_{-0.20}$ & $21.40^{+0.22}_{-0.20}$ & $21.31^{+0.22}_{-0.19}$ & $21.47^{+0.19}_{-0.20}$ \\
            \hline
		$x_1$ & Fraction of radiation processed by high-pressure clouds & $0.40^{+0.22}_{-0.17}$ & $0.40^{+0.21}_{-0.16}$ & $0.32^{+0.10}_{-0.14}$  & $0.35^{+0.22}_{-0.15}$\\
		$y_1$ & Asymmetry of high-pressure cloud distribution & $0.52^{+0.28}_{-0.23}$ & $0.53^{+0.26}_{-0.22}$ & $0.56^{+0.26}_{-0.23}$ & $0.49^{+0.28}_{-0.22}$\\
		$x_2$ & Fraction of radiation processed by low-pressure clouds & $0.20^{+0.17}_{-0.10}$ & $0.19^{+0.16}_{-0.09}$ & $0.33^{+0.21}_{-0.15}$ & $0.16^{+0.15}_{-0.08}$\\
		$y_2$ & Asymmetry of low-pressure cloud distribution & $0.40^{+0.31}_{-0.20}$ & $0.43^{+0.31}_{-0.21}$ & $0.58^{+0.26}_{-0.25}$ & $0.43^{+0.29}_{-0.21}$ \\
		$z_1$ & Line-of-sight obscuration fraction by high-pressure clouds & $0.22^{+0.20}_{-0.15}$ & $0.21^{+0.20}_{-0.14}$ & $0.21^{+0.19}_{-0.15}$ & $0.19^{+0.19}_{-0.14}$\\
		$z_2$ & Line-of-sight obscuration fraction by low-pressure clouds & $0.21^{+0.19}_{-0.14}$ & $0.20^{+0.19}_{-0.14}$ & $0.21^{+0.19}_{-0.15}$ & $0.19^{+0.20}_{-0.13}$\\
		$1-x_1-x_2$ & Isotropic escape fraction of ionizing radiation & $0.35^{+0.23}_{-0.21}$ & $0.35^{+0.22}_{-0.21}$ & $0.30^{+0.20}_{-0.19}$ & $0.43^{+0.21}_{-0.23}$\\
		$1-z_1-z_2$ & Line-of-sight escape fraction of ionizing radiation & $0.52^{+0.20}_{-0.18}$ & $0.53^{+0.19}_{-0.18}$ & $0.53^{+0.19}_{-0.17}$ & $0.56^{+0.20}_{-0.20}$\\
        \hline
        $\log({\rm C/O})$ & Carbon abundance & $-0.51^{+0.05}_{-0.05}$ & $-0.51^{+0.05}_{-0.05}$ & $-0.51^{+0.05}_{-0.05}$ & $-0.51^{+0.05}_{-0.05}$ \\
        $\log({\rm N/O})_1$ & Nitrogen abundance in high-pressure clouds & $-0.21^{+0.10}_{-0.11}$ & $-0.25^{+0.09}_{-0.11}$ & $-0.21^{+0.08}_{-0.11}$ & $-0.23^{+0.08}_{-0.11}$ \\
        $\log({\rm N/O})_2$ & Nitrogen abundance in low-pressure clouds & $\lesssim-1.31$ &$\lesssim-1.26$ & $\lesssim-1.20$ & $\lesssim-1.31$ \\
        $\log({\rm Ne/O})$ & Neon abundance & $-0.81^{+0.05}_{-0.05}$ & $-0.81^{+0.05}_{-0.05}$ & $-0.79^{+0.05}_{-0.06}$ & $-0.83^{+0.05}_{-0.05}$ \\
        $\log({\rm Si/O})$ & Silicon abundance &  $-1.84^{+0.07}_{-0.07}$ & $-1.84^{+0.07}_{-0.08}$ & $-1.84^{+0.06}_{-0.06}$ & $-1.83^{+0.07}_{-0.08}$\\
        \hline
        $\mu_{10,1}/\mu_{9,1}$ & Magnification ratio for central component & $3.30^{+0.080}_{-0.08}$& $3.29^{+0.08}_{-0.08}$ & $3.40^{+0.34}_{-0.34}$ & $3.29^{+0.08}_{-0.08}$ \\
        $\mu_{10,2}/\mu_{9,2}$ & Magnification ratio for extended component & $4.96^{+0.39}_{-0.38}$ & $4.97^{+0.38}_{-0.37}$ & --- & $4.88^{+0.42}_{-0.39}$ \\
		\hline
        \hline
	\end{tabular}
	\tablenotetext{a}{Image 10 central and extended component photometry is excluded; a single magnification ratio $\mu_{10}/\mu_9$ is applied for both components.}
    \tablenotetext{b}{Where $\log Z_{\odot} = -1.85$}
\end{table*}

\section{Discussion}
\label{sec:disc}

In this section, we discuss the physical implications of our inferred model for the cluster and the surrounding nebula. 

\subsection{Nebula Size}
\label{sec:nebsize}

The typical distance between the clouds from the cluster center can be computed from the ionizing flux striking the cloud surface $\Phi$ and the total ionizing output $Q({\rm H}^0)$. Up to a dependence on the uncertain lensing magnification factor $\mu_9$, we find that the high-pressure clouds form an inner nebula very close to the cluster at a distance $R_1 = (\mu_9/10)^{-1/2}\,(6$--$12)\,$pc, while low-pressure clouds form an outer nebula at a much larger distance $R_2 = (\mu_9/10)^{-1/2}\,(130$--$320)\,$pc. The angular diameter distance to $z_s=2.369$ is $d_{\rm A} = 1.7\,$Gpc, corresponding to an apparent proper distance $8.4\,$kpc for one arcsecond. The high-pressure nebula has an angular half-width along the direction of arc elongation, for Image 9:
\begin{align}
    \theta_{1,9} = \frac{\mu_9\,R_1}{\mu_r\,d_{\rm A}} \approx 0.006\arcsec\,\left( \frac{R_1\sqrt{\mu_9}}{10^{20.0}\,{\rm cm}} \right)\,\left( \frac{\mu_9}{10} \right)^{1/2}\,\left( \frac{\mu_r}{2} \right)^{-1},
\end{align}
where $\mu_r$ is the radial magnification. The angular half-width for the nebula comprised of the low-pressure clouds at larger distances is
\begin{align}
    \theta_{2,9} = \frac{\mu_9\,R_2}{\mu_r\,d_{\rm A}} \approx 0.15\arcsec\,\left( \frac{R_2\sqrt{\mu_9}}{10^{21.4}\,{\rm cm}} \right)\,\left( \frac{\mu_9}{10} \right)^{1/2}\,\left( \frac{\mu_r}{2} \right)^{-1}.
\end{align}
The fiducial values $\log(\sqrt{\mu_9}\,R_1)=20.0$ and $\log(\sqrt{\mu_9}\,R_2)=21.4$ we shall use are median values from the posterior distributions (c.f. \reffig{distcorner}). 

Absolute magnification factors of the various lensed images are uncertain, as existing lens models reproduce image locations well but not magnification ratios \citep{Sharon2022SunburstLensModel}. There is also uncertainty for $\mu_r$, whose value should vary moderately between lensed images. While \cite{Vanzella2022SunburstEfficiency} predicted $\mu_r=1.2\mbox{--}1.3$, \cite{Diego2022godzilla} presented two models with $\mu_r=1.4$ and $\mu_r=2.1$, respectively. The source-plane reconstruction results of \cite{Sharon2022SunburstLensModel} imply $\mu_r=1.6\mbox{--}2$. We use a high fiducial value $\mu_r=2$.

If the low-pressure nebula has a largely isotropic distribution around the cluster, the elongated image half-light full-width $\theta_{2,9}=0.15\arcsec$ is smaller than the FWHM of the PSF we use $\approx 0.27\arcsec$ in the F126N filter from which we extract a flux enhanced by {\rm [O II]}$\lambda\lambda$3726,3729. The angular half-light full-width for Image 10 is about a factor of $\approx 4$ larger, reaching $\theta_{2, 10}\approx 0.60\arcsec$, comparable to the seeing-limited FWHM $\sim 0.5\mbox{--}0.6\arcsec$ of X-shooter spectroscopy data.

Evidence for the spatially larger, low-pressure nebula is visible in Image 10 in a few filters enhanced by emission lines. The best example is the narrow filter F167N, whose flux is strongly boosted by {\rm [O III]}$\lambda$4958. The posterior distribution of this line flux shows significant contribution from both inner and outer populations of clouds. In F167N, a compact component appears superimposed on a more extended component slightly offset to one side along the arc direction. Given that our spectral model predicts much smaller stellar and nebular continuum compared to {\rm [O III]}$\lambda$4958 in F167N (c.f. \reffig{lineflux}), they probably correspond to the {\rm [O III]}$\lambda$4958 emission from the high-pressure and the low-pressure nebulae, respectively. 

If we focus on the central compact stellar component, and assumes that it is enclosed within $R_1$, the mean stellar surface density is
\begin{align}
\Sigma_\star = \frac{M_{\star,1}}{\pi\,R^2_1} = 1.9\times 10^5\,\frac{M_\odot}{{\rm pc}^2}\,\left( \frac{\mu_9\,M_{\star,1}}{10^{8.8}\,M_\odot} \right)\,\left( \frac{R_1\,\sqrt{\mu_9}} {10^{20.0}\,{\rm cm}} \right)^{-2},
\end{align}
independent of the uncertain magnification $\mu_9$. This is on the order of the maximum surface density $\Sigma_\star \simeq 10^5\,M_\odot\,{\rm pc}^{-2}$ observed in local Universe stellar systems \citep{Hopkins2010MaximumSurfaceDensity, KMBH2019ARAAreview}, but a top-heavy IMF can lower this value.

The line-of-sight cloud velocity corresponding to a radius $R_1$ is
\begin{align}
\label{eq:vR1}
v(R_1) = & \left(\frac{G\,M_{\star,1}}{3\,R_1}\right)^{\frac12} \approx 90\,{\rm km}\,{\rm s}^{-1}\,\left( \frac{\mu_9\,M_{\star,1}}{10^{8.8}\,M_\odot} \right)^{1/2}\nonumber\\
& \times \left( \frac{R_1\,\sqrt{\mu_9}} {10^{20.0}\,{\rm cm}} \right)^{-1}\,\left(\frac{\mu_9}{10}\right)^{-\frac14},
\end{align}
where we exclude the extended stellar component as its size is significantly beyond $R_1$. { This is larger than the velocity dispersion of Milky Way globular clusters, but our inferred stellar mass $M_{\star,1}$, for a probably magnification factor $\mu_9=10$, is also more than an order of magnitude larger than that of the present-day Milky Way globular clusters}. On the other hand, this is larger than the observed $40\,{\rm km}\,{\rm s}^{-1}$ quoted by \cite{Vanzella2022SunburstEfficiency}. Adopting a more top-heavy IMF will lead to a smaller $\mu_9\,M_{\star,1}$, and hence a smaller $v(R_1)$, which may mitigate the discrepancy.

\begin{figure}[t]
    \centering
    \includegraphics[width=8.5cm,clip]{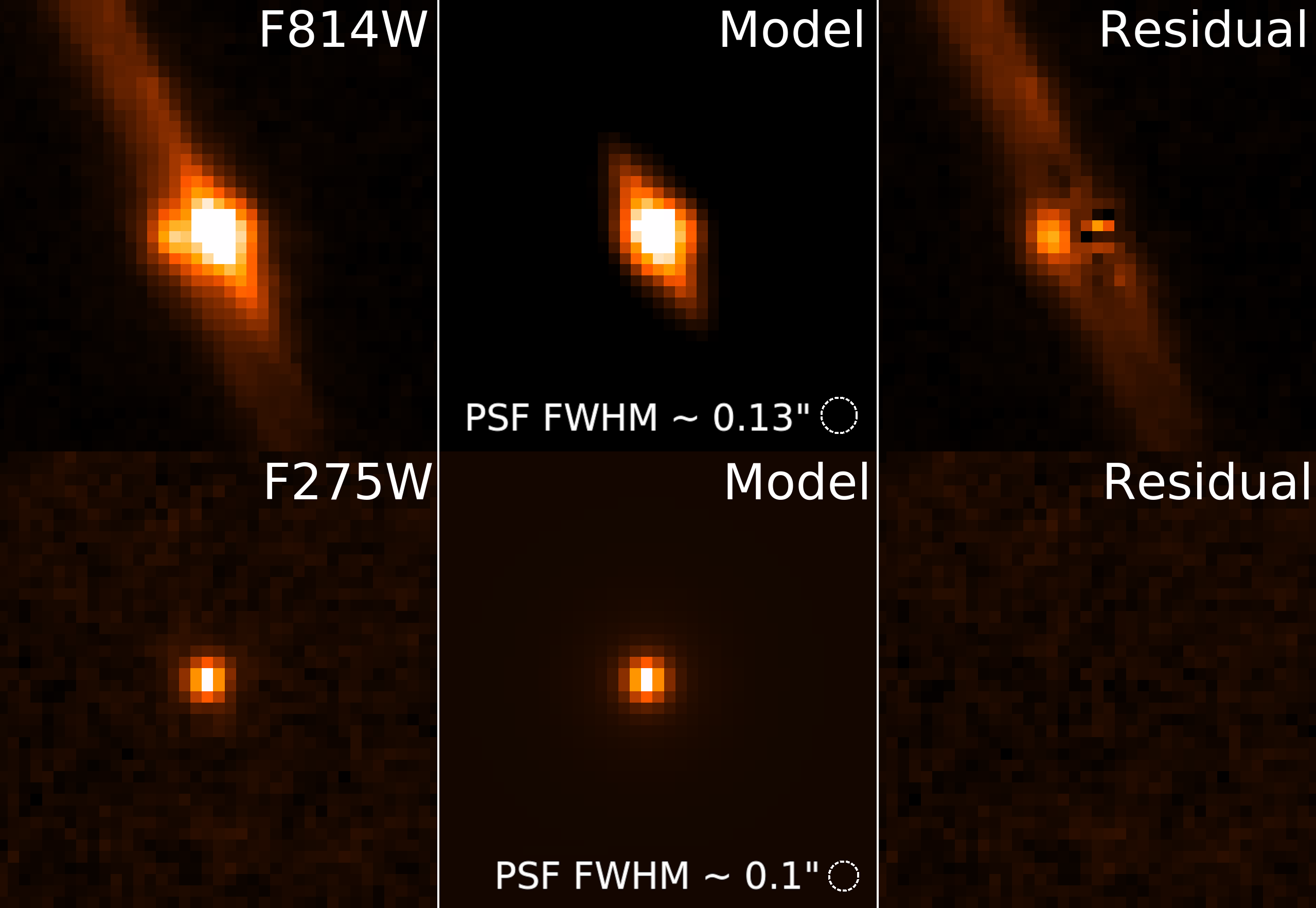}
     \caption{Spatially resolved modeling of Image 10. In the top panels, continuum image in F814W shows that stars in the LyC cluster can be modeled as the sum of a central unresolved component and an extended component modeled as a 1D Gaussian profile (due to lensing shear). The 1D Gaussian is centered at an angular separation $0.015\arcsec \pm 0.001\arcsec$ away from the central component, with a Gaussian width $\sigma=0.15\arcsec \pm 0.01\arcsec$. However, in the F275W filter which probes escaping LyC radiation, the extended component is not visible. This implies that only LyC photons emitted from the central compact stellar component manage to escape.}
    \label{fig:astrometric}
\end{figure}

\subsection{Nebula Mass and Column Density}
\label{sec:nebcolmass}

Depending on the column density, the high-pressure clouds amount to mass
\begin{align}
\label{eq:Mneb2}
    M_c \simeq & 1.4 \times 4\pi\,R^2_1\,N_{\rm H}\,m_p\,x_1 \nonumber\\
    = & 6\times 10^5\,M_\odot\,\left( \frac{N_{\rm H}}{10^{23}\,{\rm cm}^{-2}} \right)\,\left( \frac{\sqrt{\mu_9} R_1}{10^{20.0}\,{\rm cm}} \right)^2  \nonumber\\
    & \times \left( \frac{\mu_9}{10} \right)^{-1}\,\left( \frac{x_1}{0.4} \right),
\end{align}
For a comparison, the cumulative stellar mass loss {through massive star winds from the central stellar component, as predicted by the best-fit BPASS model up to} $t_{\rm age} = (3\mbox{--}4)\,$Myr, is
\begin{align}
\label{eq:delMcl}
    \Delta M_\star = (1\mbox{--}2) \times 10^6\,M_\odot\,\left( \frac{\mu_9\,M_{\star,1}}{10^{8.8}\,M_\odot} \right)\,\left(\frac{\mu_9}{10}\right)^{-1}.
\end{align}
This is $2$--$4\%$ of the initial stellar mass assuming the standard IMF.

We attempted to constrain the H I column $N_{\rm HI}$ from data by fixing cluster and nebula intensive parameters to the best-fit values but allowing extensive and geometric parameters to freely vary. Additionally, we varied $\log N_{\rm HI}\,[{\rm cm}^{-2}]$ from 21 to 24 in \texttt{Cloudy} calculations. We were however unable to obtain any significant constraint on $N_{\rm HI}$, mainly because no emission lines that specifically probe the cool H I gas are available, and because dust attenuation through the H I column is strongly degenerate with extensive and geometric parameters. UV absorption lines are not useful either because the dusty clouds severely attenuate any UV light emerging from the shaded faces.

Nevertheless, dynamical consideration constrains $N_{\rm HI}$ for the high-pressure clouds. {From the inferred ionizing photon flux striking the cloud surfaces $\Phi({\rm H_0})$ for the high-pressure clouds}, the H II gas has a thickness $\sim 0.02\,{\rm pc}/(\nH/10^5\,{\rm cm}^{-3})$ and a column $N_{\rm HII} \sim 6 \times 10^{21}\,{\rm cm}^{-2}/(\nH/10^5\,{\rm cm}^{-3})$. If the H II gas is density bounded, $\nH \sim 10^5\,{\rm cm}^{-3}$ leads to a sufficient radiation pressure coupling with LyC photons that would rapidly eject the clouds out of the cluster potential compared to the free-fall timescale
\begin{align}
    t_{\rm ff} & = \frac{\pi}{2}\,\left( \frac{R^3_1}{G\,M_{\star,1}} \right)^{\frac12} \nonumber\\
    & \approx 10^5\,{\rm yr}\,\left( \frac{\mu_9\,M_{\star,1}}{10^{8.8}\,M_\odot} \right)^{-\frac12}\,\left( \frac{R_1\,\sqrt{\mu_9}} {10^{20.0}\,{\rm cm}} \right)^{\frac32}\,\left(\frac{\mu_9}{10}\right)^{-\frac14}.
\end{align}

If instead the H II gas is ionization bounded, a minimum column $\log N_{\rm HI} > 22.8$ is required for cluster gravity to counteract radiation pressure and retain the cloud \citep{Fall2010StellarFeedbackYSC, Murray2010RadiationFeedbackGMC, KMBH2019ARAAreview}. For even larger $N_{\rm HI}$, the bulk of the cloud would fall back to the cluster, but no faster than $t_{\rm ff}$. For $23 < \log N_{\rm HI} < 25$, dust reprocessed IR radiation trapped in the H I gas may drive a slow outflow \citep{Menon2022IRFeedbackII}, albeit the outflow velocity may be insignificant given the large escape velocity $v_{\rm esc} \sim 100\,{\rm km}\,{\rm s}^{-1}$ involved here. Seeing short-lived clouds with a lifetime $< 10^5\,$yr would appear unnatural given $t_{\rm age}\lesssim4\,$Myr, unless these clouds are continuously replenished.

The dusty cloud interior should cool down to a low temperature typical of molecular clouds, with an increased density $\sim 10^6\mbox{--}10^7\,{\rm cm}^{-3}$ as a result of pressure equilibrium. The high-pressure clouds have a typical thickness
\begin{align}
\label{eq:l}
    l = 0.03\,{\rm pc}\,\left( \frac{N_{\rm H}}{10^{23}\,{\rm cm}^{-2}} \right)\,\left( \frac{\nH}{10^6\,{\rm cm}^{-3}} \right)^{-1}.
\end{align}
Since the dusty interior can cool down to very low temperatures, the pressure support in the cloud interior is likely dominated by turbulence \citep{McKeeTan2003TurbulentCores}. In this case, \refeq{l} holds whether or not the interior is atomic or molecular.
Clouds are out of virial equilibrium and subject to collapse for pressures $P\la 3\pi\,G\,\Sigma^2/20$ \citep{KMBH2019ARAAreview}. Correspondingly, if the high-pressure clouds have a column $N_{\rm H} \gtrsim 0.9 \times 10^{24}\,{\rm cm}^{-2}\,(P_1 / 10^9\,{\rm K}\,{\rm cm}^{-3})^{1/2}$  and survive long enough near the star cluster, they can form stars internally. At this value of $N_{\rm H}$, the total cloud mass may be too large to come entirely from cluster mass loss (c.f. \refeq{delMcl}), and the clouds would have to be a mixture of stellar ejecta and leftover natal gas.

A delicate balance between outward radiation pressure and inward gravity may be unrealistic. If a cloud migrates to a different radius from the cluster center, the characteristic ambient pressure will change, which will in turn result in a pressure change in the cloud neutral interior. Assume that the ambient pressure scales with cluster-centric radius $r$ as $P \propto r^{-\alpha}$ with $\alpha>0$. If this pressure scale is set by radiation pressure, for instance, we may have $\alpha=2$. If the dust-shielded, cold cloud interior is supported by either thermal pressure at a fixed temperature, or by turbulence at roughly a fixed velocity dispersion, the particle number density will scale as $n \propto r^{-\alpha}$, and the mass column density will scale as $\Sigma \propto r^{-2\,\alpha/3}$ under the assumption that the cloud geometry maintains an order-unity aspect ratio. This means that $P/\Sigma^2 \propto r^{\alpha/3}$. If gravity overcomes radiation pressure and pulls a cloud to a smaller $r$, it further pressurizes and the threshold for Jeans instability is easier to reach. Conversely, if radiation pressure overcomes gravity and pushes the cloud outward, the cloud will expand and will never form stars. If the observed high-pressure clouds can survive other possible mechanisms of shattering or destruction and sink toward the cluster center due to its weight, they may form stars internally.

When the star cluster is mass segregated, the most intensely radiating stars cluster in the core. In this situation, radiation pressure increases more steeply than gravity at distances smaller than the cluster's outer edge. We note that this may allow an intermediate range of distances where the balance between radiation pressure and gravity could be stable for clouds of suitable column density.

For a realistic cloud geometry radiation likely manages to expel peripheral parts of the clouds along low column density sight-lines even if the cloud bulk is retained by gravity \citep{YehMatzner2012RadiationWindFeedback, ThompsonKrumholz2016RadiationPressure, Raskutti2017RadiationGasInteraction}. Without replenishment, the clouds are not expected to survive for many millions of years. It would be interesting to observationally determine $N_{\rm H}$ in order to better understand the dynamical origin and fate of the pressurized dusty clouds. 

\subsection{Nebula Pressure}
\label{sec:nebpressure}

For the high-pressure clouds, the ionizing flux $\log\Phi_1({\rm H}^0) \approx 14.3^{+0.3}_{-0.2}$ imparts a radiation pressure $\log P_{1,{\rm rad}}=9.0\mbox{--}9.5$. Further including dust absorption of UV light increases the total radiation pressure by about three fold to $\log P_{1,{\rm rad}}=9.5\mbox{--}10.0$. Compression by radiation appears to be a viable dynamical origin for the high-pressure clouds within our inference uncertainty. For our best-fit incident SED, external radiation pressure equals H II gas pressure if $\log U_1 = -1.4$. From C III]$\lambda\lambda$1908,1906 line ratio, the H II density $\log\nele \approx 5$ \citep{Kewley2019NebulaPressure} at $T_{\rm e} = 15000\,$K corresponds to a gas pressure $\log P_{1,{\rm gas}} \approx 9.5$. The high-pressure clouds are in the regime where external radiation pressure is comparable or slightly higher than the H II gas pressure.

For the low-pressure clouds, on the other hand, the ionizing flux $\log\Phi_2({\rm H}^0) \approx 11.5^{+0.3}_{-0.3}$ converts into a radiation pressure $\log P_{2,{\rm rad}} \approx 6.6$--$7.2$, which is subdominant than the inferred total pressure $\log P_2$. Radiation pressure therefore appears dynamically unimportant for the low-pressure clouds.

Another possible source of confining pressure is kinetic feedback from stellar winds and SN explosions \citep{KooMcKee1992WindBubblesI, KooMcKee1992WindBubblesII, RogersPittard2013WindSNFeedback, Calura2015WindFeedback, Wareing2017MechFeedback}, and other means of cluster mass loss \citep{deMink2009BinaryAbundanceAnomaly}. Taking $Z=0.2\,Z_\odot$ and at $t_{\rm age}=2\mbox{--}4\,$Myr, the BPASS (v2.2) model predicts a total mass loss rate
\begin{align}
\label{eq:mdot}
    \dot M_{\star,1} \approx 1\,M_\odot\,{\rm yr}^{-1}\,\left( \frac{\mu_9\,M_{\star,1}}{10^{8.8}\,M_\odot} \right)\,\left(\frac{\mu_9}{10}\right)^{-1},
\end{align}
and a total mechanical luminosity dominated by the first SN explosions (onset at $t_{\rm age} \approx 3\,$Myr)
\begin{align}
\label{eq:lmech}
    L_{\rm mech} \approx 6\times 10^{41}\,{\rm erg}\,{\rm s}^{-1}\,\left( \frac{\mu_9\,M_{\star,1}}{10^{8.8}\,M_\odot} \right)\,\left(\frac{\mu_9}{10}\right)^{-1}.
\end{align}
If the first SNe have not yet exploded, pure wind contributions to $L_{\rm mech}$ would be a factor of $\sim 5$ smaller. If these mass loss thoroughly thermalizes to drive a supersonic, non-radiative cluster wind \citep{Chevalier1985Nature, CantoRagaRodriguez2000ClusterWind, Wunsch2011ClusterWindWithCooling, Silich2011ClusterWindExponentialProfile}, the wind reaches a velocity $v_{\rm w}=(2\,L_{\rm mech}/\dot M_{\star,1})^{1/2} \approx 1400\,{\rm km}\,{\rm s}^{-1}$, with uncertainty on $\mu_9$ and $M_{\star, 1}$ canceling out. If the cluster wind is fully blown well within radius $R_1$, it should exert a ram pressure on the high-pressure clouds:
\begin{align}
    P_{\rm ram} & = \frac{\dot M_{\star,1}\,v_{\rm w}}{4\pi\,R^2_1} =  3.8 \times 10^9\,{\rm K}\,{\rm cm}^{-3}\,\left( \frac{\mu_9\,M_{\star,1}}{10^{8.8}\,M_\odot} \right)\nonumber\\
    & \times \left( \frac{R_1\,\sqrt{\mu_9}} {10^{20.0}\,{\rm cm}} \right)^{-2}\,\left( \frac{v_{\rm w}}{1400\,{\rm km}\,{\rm s}^{-1}} \right).
\end{align}
This is comparable to the inferred total pressure $P_1$ or the expected radiation pressure. Theoretical studies have suggested that for a very compact, massive star cluster the cluster wind may be subject to efficient or even catastrophic radiative cooling \citep{Silich2004RadiativeClusterWindS, Wunsch2011ClusterWindWithCooling, Gray2019SuperwindCoolingI, Danehkar2021SuperwindCoolingII, LochhaasThompson2017Cooling, Wunsch2017RapidCoolingRHDsimulation}, which can further reduce kinetic feedback, {which we investigate further in the next sub-section.}

\begin{figure}[t]
    \centering
    \includegraphics[width=\columnwidth]{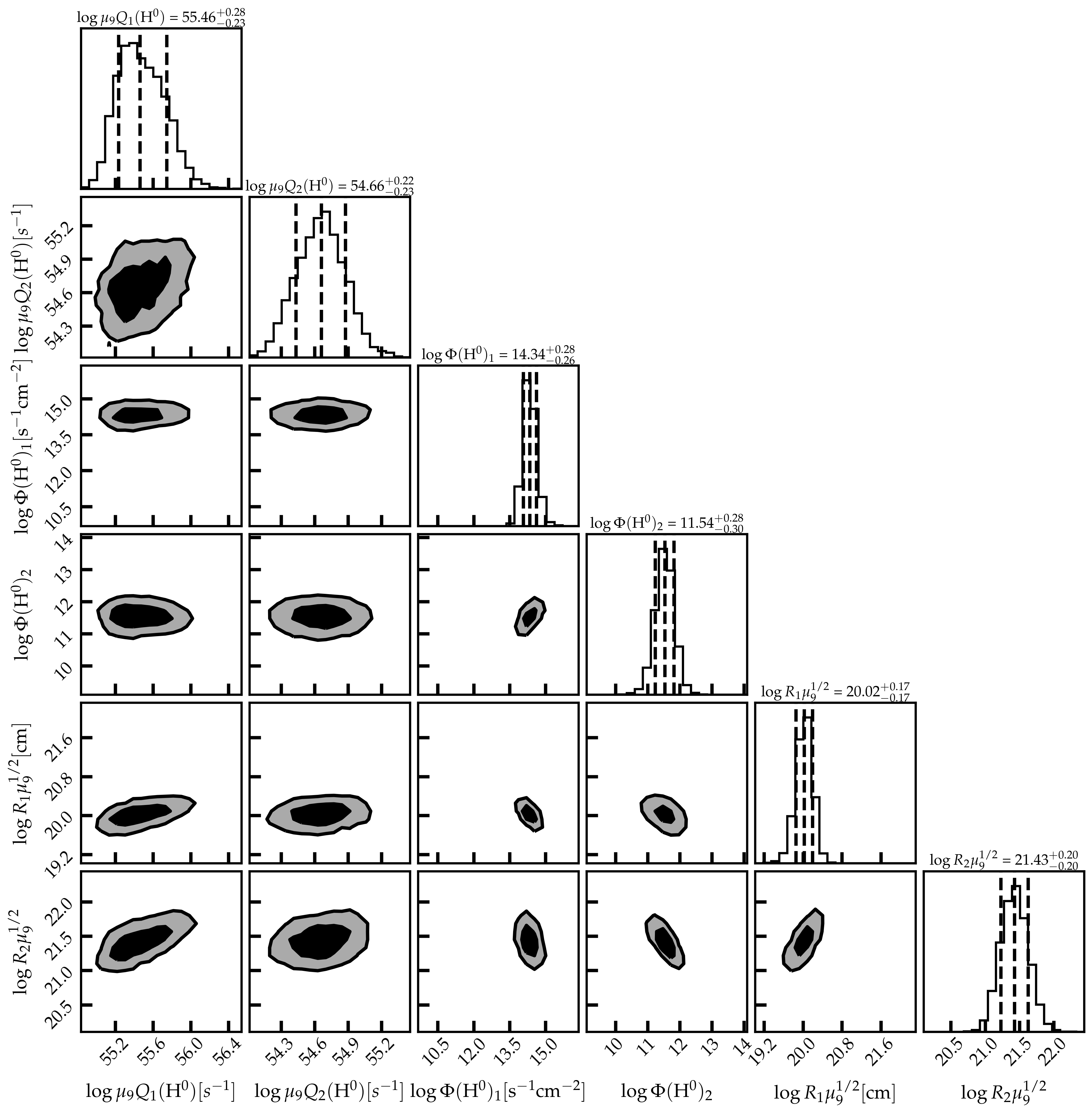}
    \caption{Posterior distributions for the measured ionizing photon rate, ionizing flux, and the typical distance from the cluster center for each cloud population. Contours levels and histogram confidence intervals follow Fig.~\ref{fig:starscorner}.
    Refer to the ``Two-Comp'' model in Table~\ref{tab:cloudtable} for the 68\% C.I.'s of the parameters.
    }
    \label{fig:distcorner}
\end{figure}

\subsection{Nitrogen Enrichment}
\label{sec:nitrogen}

\begin{figure}[t]
    \centering
    \includegraphics[width=8.25cm,trim=0.3cm 0.25cm 0.25cm 0.25cm,clip]{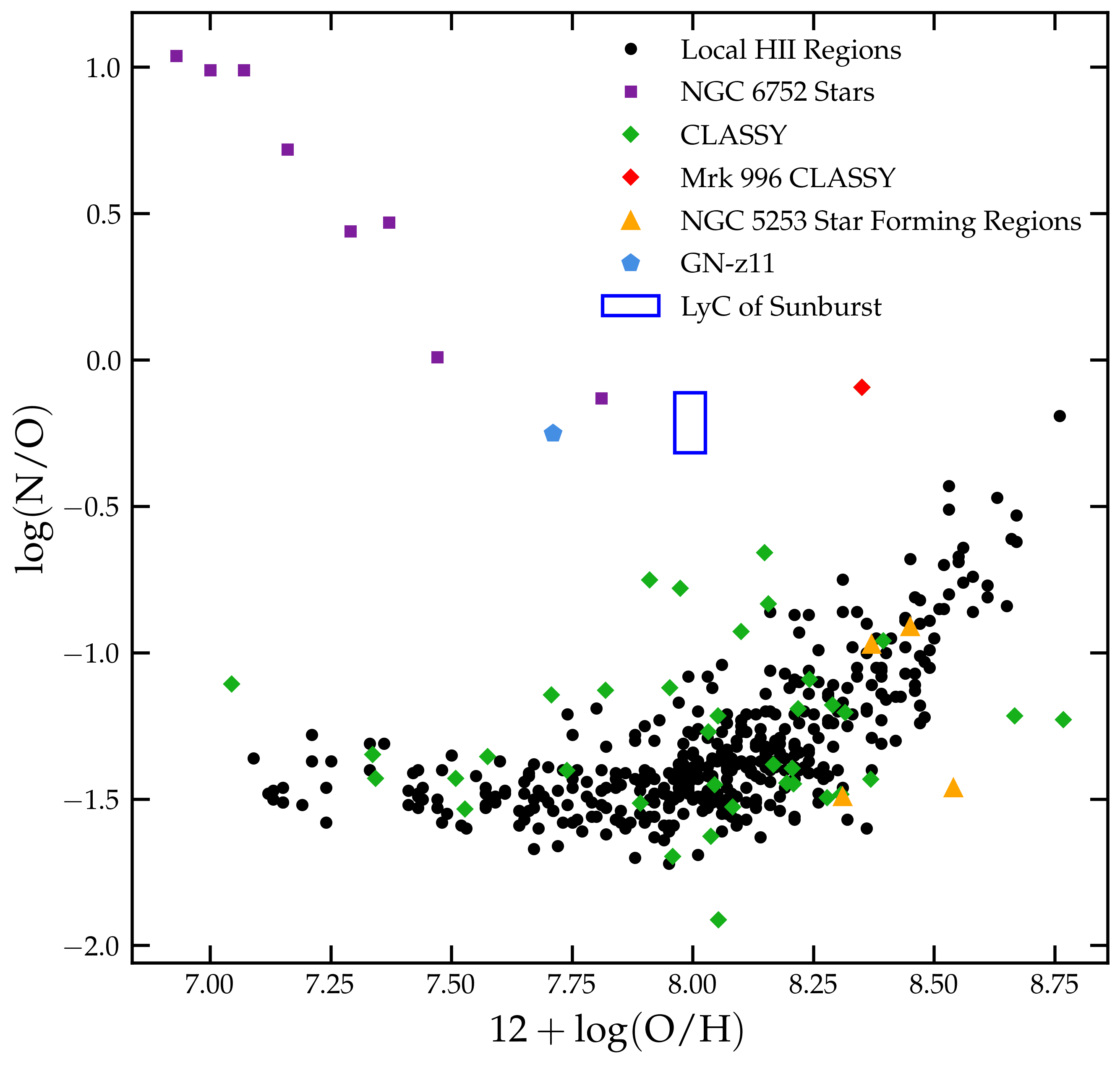}
    \caption{Nitrogen abundances of the LyC cluster compared to post-turnoff dwarf stars of local globular clusters \cite[NGC 6752][]{Carretta2005NGC6752}, star forming regions of NGC 5253 \citep{LopezSanchez2007NGC5253}, local star forming galaxies \cite[CLASSY]{Berg2022CLASSY,Stephenson2023CLASSY}, and local HII regions in SDSS galaxies (\cite{Pilyugin2012HIIRegions}). Our photoionization calculation infers a nitrogen abundance comparable to highest seen in local HII and star forming galaxies. Our measured abundance pattern falls on a diagonal line (which goes from upper left to lower right) that is traced by abundances measured from chemically anomalous dwarf stars in the Milky-Way globular cluster NGC 6752, which indicates the various degree of mixing between N-enriched gas and chemically normal gas. Finally, we also compare to GN-z11, a $z\sim11$ galaxy hosting candidate globular cluster progenitors \citep{Bunker2023gnz11,Cameron2023gnz11,Charbonnel2023gnz11,Senchyna2023gnz11}.}
    \label{fig:nitrogen}
\end{figure}

The N III] multiplet near $1750\,\AA$ \citep{FeldmanDoschek1979NiiiOiv} is not commonly reported as a strong UV line from H II nebulae as N is significantly less abundant than O and C in the ISM. Among a sample of 45 analogs of high-$z$ starburst galaxies at $0.002 < z < 0.182$ from the CLASSY survey \citep{Berg2022CLASSY}, there is only one strong detection of this multiplet from the blue compact dwarf galaxy Mrk 996, and a lower significance detection in ${\rm S/N}\sim 3$ from another system \citep{Mingozzi2022CLASSY}. The N III] detection in Mrk 996 is toward the nuclear region and indicates hugely enhanced N abundance $\log({\rm N/O})=0.0$ possibly related to ejecta from WN stars. However, a large velocity dispersion $\sim 450\,{\rm km}\,{\rm s}^{-1}$ is associated with the N III] emitter \citep{Thuan1996Mrk996}, and it is unclear whether the example in Mrk 996 shares a similar origin from nebulae around a super star cluster as in Sunburst. On the other hand, a three-fold N enrichment $\log({\rm N/O})=-0.85$ localized to a few super star clusters is reported in NGC 5253 \citep{Kobulnicky1997NGC5253, LopezSanchez2007NGC5253}.

The detection of {\rm N III]}$\lambda\lambda$1750,1752 in this paper, with $\log({\rm N/O})= -0.21^{+0.10}_{-0.11}$, is localized to the LyC cluster vicinity but unseen in other parts of the arc. \cite{Mestric2023VMS} mentioned the detection of {\rm N III]}$\lambda$1750 from the LyC cluster without giving further comments. The facts that (1) the $\lambda\lambda$1750,1752 lines show as the strongest among all five transitions of the N III] multiplet and (2) the observed line ratio $\lambda$1750/$\lambda$1752 $\approx 2$ are consistent with the inferred density of the high-pressure clouds. This indicates an N abundance that is increased by $\sim 12$--$19$ fold relative to the median interstellar N abundance $\log({\rm N/O})\approx-1.4$ at $Z \approx 0.2-0.3\,Z_\odot$ \citep{Molla2007NitrogenEvolution, Berg2019MetalPoorCNOEvolution} in the very vicinity of the star cluster, a scenario previously theorized by \cite{Izotov2006logZemislingalaxies}. This level of N enrichment is higher than what is found near several super star clusters in NGC 5253 \citep{KobulnickyKewley2004}, and the N/O ratio is higher than what is observed from extragalactic H II regions in low metallicity emission (see Fig. 11 of \cite{Izotov2006logZemislingalaxies}). In contrast to the $z\approx 0$ HII regions, the nitrogen enhancement we find in the nebular gas is comparable to that seen in post-turnoff dwarf stars of Milky-Way's globular clusters such as NGC 6752 \citep{Carretta2005NGC6752}.

Shortly after the completion of this work, four pre-prints have appeared and have examined remarkably strong nitrogen emission lines seen in GN-z11, a $z \approx 11$ galaxy confirmed through JADES NIRSpec IFU spectroscopy \citep{Bunker2023gnz11,Cameron2023gnz11,Senchyna2023gnz11,Charbonnel2023gnz11}. Using \texttt{Cloudy} photoionization calculations, \cite{Senchyna2023gnz11} showcase a similarly high density $n_e \gtrsim 10^5\,{\rm cm}^{-3}$ for photoionized gas, and a high nebular N abundance $\log({\rm N/O}) \sim -0.25$ at a gas metallicity of $Z \approx 0.1 Z_{\odot}$. This object may be a high-redshift, high-mass (stellar mass $\sim 10^8$--$10^9\,M_\odot$ \cite{Bunker2023gnz11}) version of the enrichment processes seen in Sunburst. \reffig{nitrogen} shows the inferred LyC cluster nitrogen abundance compared to GN-z11, as well as to individual chemically anomalous stars in a local globular cluster, local star forming galaxies, and local HII regions.

We surmise that heavy N enrichment might be typical in the vicinity of very young super star clusters, and is indicative of either pollution of natal gas by WN stars or LBVs, or condensation solely from the wind material of those stars, or even ejecta from interacting binary massive stars \citep{deMink2009BinaryAbundanceAnomaly}, which happens as early as $\sim 3\mbox{--}4\,$Myr after the onset of star formation. 
The CNO cycle $^{14}{\rm N}({\rm p},\gamma)^{15}{\rm O}$ bottleneck causes efficient conversion of C and O to N, such that ejecta mixed up with H-burning-processed material can be significantly N enhanced but is expected to have a conserved sum of C, N and O. Given the solar C:N:O ratio of 4:1:7, complete conversion of C and O into N can cause up to 12-fold increase in the absolute abundance of N in ejected material. Since the observed photoionized gas {does not show extreme depletion of C and O abundances relative to elements like Ne}, the CNO-processed material must mix with other chemically normal gas. The observed high ${\rm N/O} \sim 0.5$--$0.8$ indicates that the CNO-processed material accounts for an order-unity mass fraction of the dense gas (also see Figure 1 of \cite{Charbonnel2023gnz11}). Since we have estimated that the mass of the dense, N-enriched gas is substantial compared to the total massive star mass loss (\refeq{Mneb2}) based on the arguments of column density and efficient turbulence mixing (see the remainder of this sub-section), we suggest that high N abundance is evidence for radiation-hydrodynamic processes that have caused massive star ejecta to efficiently condense and be retained in the cluster potential.

High-pressure clouds condensed solely out of cooled SN ejecta is not favored since the high-pressure clouds do not appear rich in oxygen and carbon. The inferred cluster age constraint $t_{\rm age}\lesssim4\,$Myr makes it unclear whether the first SNe have exploded {(perhaps up to this age all massive stars have collapsed to black holes without ejecting significant amount of O, Mg, Si and Fe as SNe with lower-mass progenitors do)}, and if SNe do have exploded, whether the SN ejecta remains hot and quickly leaks out of the cluster vicinity via low density channels \citep{Lopez2011R136Feedback, RogersPittard2013WindSNFeedback, RogersPittard2014WindSNFeedbackII}. At radius $R_1$, the escape velocity out of the cluster's potential $v_{\rm esc}(R_1) \sim 220\,{\rm km}\,{\rm s}^{-1}$ (a factor of $\sqrt{6}$ larger than \refeq{vR1}) is slow compared to that of the typical O-star/WR-star wind or a fully blown, hot cluster superwind. { However, rapid radiative cooling may provide a route for retention of these ejecta. We follow the semi-analytic approach of \cite{LochhaasThompson2017Cooling} to assess whether the cluster is in a regime where radiative cooling will retain rejecta within the cluster radius $R_1$. We assume the mass loss rates and mechanical luminosities given in \refeq{mdot} and \refeq{lmech} respectively, which we find are consistent with the linear scalings used by \cite{LochhaasThompson2017Cooling}. We use the same cooling function for $T>10^5\, {\rm K}$ from \cite{Draine2011HIIRegionRadiationPressure} as used in \cite{LochhaasThompson2017Cooling}, but, following \cite{MaclowMcCray1988}, scale the cooling rate linearly with metallicity such that $\Lambda_0 \simeq 4.4 \times 10^{-23}\,{\rm erg }\,{\rm s}^{-1}\,{\rm cm}^3$ for $Z \simeq 0.2Z_{\odot}$. Given these assumptions and assuming a thermalization efficiency $\alpha \simeq 0.1$ for merged stellar winds from individual stars and a conservative mass-loading factor of order unity (see discussion earlier in this subsection), we solve Eq.(14) in \cite{LochhaasThompson2017Cooling} and find that it is possible for ejecta from hot massive star winds to cool and be retained within a radius $R_1$ for the LyC star cluster. However, given the likely small mass-loading factor of the nebula, it cannot be ruled out that a larger thermalization efficiency or a larger magnification factor could push the system out of the regime of catastrophic radiative cooling.}

{Slow outflows from massive rotating stars may also provide a suitable enrichment mechanism. This may be in the form of a single massive star rotating near its critical velocity, where rotational mixing causes dredge-up of heavy elements from the inner convective zone and allows for envelope material to be liberated in a slow wind \citep{Roy2022PreSNwindsNitrogen}.}

{Finally, ejecta from non-conservative binary mass transfer may also explain the observed Nitrogen enrichment \citep{deMink2009BinaryAbundanceAnomaly}. A massive star may shed nearly half of the mass in interaction with a lower mass companion. The companion may then spin up to almost the critical rotation velocity as it accretes from the donor star and stops to accept transferred material. This then causes an outflow of H-burning processed material from the inner layers of the donor star, typically with a slow outflow velocity comparable to the binary orbital velocities.}


In our best-fit model, the high-pressure clouds have a thickness given by \refeq{l}, so that turbulent motion $\sigma \approx (P_1/n_{\rm H}\,m_{\rm p})^{1/2} \approx 4\,{\rm km}\,{\rm s}^{-1}\,(n_{\rm H}/10^6\,{\rm cm}^{-3})^{-1/2}$ should quickly mix N throughout the cloud on a timescale
\begin{align}
    t_{\rm mix} \approx \frac{l}{\sigma} = 8000\,{\rm yr}\,\left( \frac{N_{\rm H}}{10^{23}\,{\rm cm}^{-2}} \right)\,\left( \frac{n_{\rm H}}{10^6\,{\rm cm}^{-3}} \right)^{-\frac12}.
\end{align}
Given the high $\log({\rm N/O})$ and the estimated total mass for the high-pressure clouds \refeq{Mneb2}, if the clouds are gravitationally retained, we estimate a total nitrogen mass under the assumption of thorough mixing
\begin{align}
\label{eq:MN}
    M_N \approx & 500\,M_\odot\,\left( \frac{N_{\rm H}}{10^{23}\,{\rm cm}^{-2}} \right)\,\left( \frac{\mu_9^{1/2} R_1}{10^{20.0}\,{\rm cm}} \right)^2  \nonumber\\
    & \times \left( \frac{\mu_9}{10} \right)^{-1}\,\left( \frac{x_1}{0.4} \right)\,\left(\frac{{\rm N/O}}{0.6}\right)\,\left(\frac{Z}{0.2\,Z_\odot}\right).
\end{align}
Until $3$\,Myr, about $8400\,(\mu_9\,M_{\star,1}/10^{8.8}\,M_\odot)\,(\mu_9/10)^{-1}$ stars more massive than $140\,M_\odot$ should have evolved to the end of their lives \citep{Bressan2012PARSEC}. By $4\,$Myr, the minimum mass becomes $60\,M_\odot$ and the number becomes $3.6\times 10^4\,(\mu_9\,M_{\star,1}/10^{8.8}\,M_\odot)\,(\mu_9/10)^{-1}$. For an average nitrogen yield $(0.01\mbox{--}0.1)\,M_\odot$ from each such star \citep{Prantzos2018RotatingMassiveStars}, the maximum nitrogen pollution should be about $\sim (84\mbox{--}3600)\,M_\odot\,(\mu_9\,M_\star/10^{8.8}\,M_\odot)\,(\mu_9/10)^{-1}$, the estimate of which is insensitive to IMF assumptions. The high-pressure clouds therefore appear to have retained an order-unity fraction of the ejected nitrogen. As we have argued before, an order-unity fraction of the cloud gas must have originated from massive star ejecta to account for the observed enhancement. Given the expected nitrogen budget, \refeq{MN} suggests that $\log N_{\rm H} \gtrsim 24$ is unlikely unless the clouds are not well mixed and are only partially polluted with nitrogen on the illuminated side which includes the H II zone. The column density of the high-pressure clouds may barely reach the threshold of secondary star formation without violating the nitrogen budget constraint, unless they pressurize further.

\section{Conclusion}
\label{sec:concl}

High SNR archival data owing to large gravitational magnifications has enabled us to closely study a newborn, LyC-leaking super star cluster in the Sunburst galaxy at Cosmic Noon. We have used the \texttt{BPASS} stellar population SED templates that account for radiation from binary stars. Using these templates as the input, we have computed the observed spectrum from the system by performing self-consistent photoionization calculations using \texttt{Cloudy}, taking into account the dependence on nebula pressure, incident radiation strength, and nebula geometry. We have constructed a grid of spectral models from these calculations and have applied them to inferring physical parameters describing the super star cluster (summarized in \reftab{cloudtable}). 

We have demonstrated that a detailed, physically motivated photoionization model can be constructed for a super star cluster surrounded by photoionized clouds, and that a joint photometric and spectroscopic analysis enables comprehensive extraction of key physical parameters despite the tremendous cosmic distance to the source.

We have determined that the LyC cluster is very massive ($M_\star \sim {\rm few} \times 10^7\,M_\odot$), $\lesssim4\,$Myr old, and has a sub-solar metallicity $Z \approx 0.2\,Z_\odot$. Little dust was detected from the large-scale ISM of the host galaxy (see \S6.1 and \reffig{starscorner}). Assuming the standard IMF, the cluster's surface stellar mass density is on the order of the maximum observed in massive stellar systems in the local Universe, $\Sigma \sim 10^5\,M_\odot\,{\rm pc}^{-2}$~\citep{Hopkins2010MaximumSurfaceDensity}.

We have shown that the cluster stars irradiate highly-pressurized ($P\sim 10^9\,{\rm K}\,{\rm cm}^{-3}$), probably density-bounded photoionized clouds { located $\lesssim 10$ pc away from the cluster center} (\S6.2, \reffig{cloudcorner}). The clouds show normal C/O and Ne/O abundance ratios but appear highly enriched with nitrogen, having $\log({\rm N/O}) \approx -0.21$ (\S6.4, \reffig{abundances}). The clouds also exhibit depleted gas-phase Si and likely have self-shielded, dusty interior. To avoid rapid ejection from the system by radiation pressure, these clouds must have large enough column densities and hence amount to at least a few $\times 10^5\,M_\odot$ of gas, and $\sim 500\,M_\odot$ of nitrogen if chemically thoroughly mixed with a typical thickness of $\sim0.03 {\rm pc}$. This requires efficiently retaining nitrogen from condensed massive star ejecta as up to 4 Myr stars heavier than $60\,M_\odot$ can yield up to $3600\,M_\odot$ of nitrogen in optimistic estimates (see \S7.4). Detection of optical {\rm [O II]} doublet which requires a low ionization condition implies that a less dense nebula extends to tens of pc or beyond, and might have already been resolved in a few emission-line dominated HST filters. We have found that nitrogen is probably not as significantly enriched in this low-density component. While in our modeling we have introduced two cloud populations with disparate properties, it might as well be that there is a continuous distribution of photo-ionized clouds, with decreasing pressure, density and ionization parameter as the cluster-centric distance increases. Through modeling the nebula emission relative to direct star light, we have inferred that $(10\mbox{--}50)\%$ of the ionizing radiation escapes the cluster's vicinity along unblocked slight lines, and that this fraction is as high as $(30\mbox{--}70)\%$ along our slight line (\S6.3, \reffig{geocorner}), consistent with previous works \citep{RiveraThorsen2019Sci}.

The origin of the high-pressure clouds probably connects to radiation-hydrodynamic evolution of massive star winds, SNe, or binary ejecta from massive stars in the cluster potential. Mass loss mechanisms that predominantly produce slowly moving ejecta ($\lesssim 300\,{\rm km}\,{\rm s}^{-1}$) are favored. The inferred element abundance is more consistent with pollution by nitrogen-enriched H-burning-processed material than oxygen-rich He-burning-processed material. Since observational selection by strong lensing is likely independent of these source properties, we suggest that this phenomenon may be commonplace in the vicinity of newborn super star clusters. We have argued that these clouds likely have a large column density $N_{\rm H} \gtrsim 10^{22.8}\,{\rm cm}^{-2}$. They might allow star formation in their interior if $N_{\rm H} \gtrsim 10^{24}\,{\rm cm}^{-2}$, or if they sink further toward the cluster in the future. It is intriguing to ponder the question whether the nitrogen enriched gas connects to the puzzle of multiple stellar populations in GCs~\citep{LochhaasThompson2017Cooling, Wunsch2017RapidCoolingRHDsimulation}, as second-generation GC stars often exhibit elevated N abundance~\citep{Bastian2018ARAAGCMultiPop, Roy2022PreSNwindsNitrogen}.

There are future observational prospects for the Sunburst galaxy to determine the physical origin of the dense photoionized clouds. Upcoming JWST imaging and spatially resolved spectroscopy (GO \#2555; PI: T. E. Rivera-Thorsen) in near-IR will complement existing HST data and will put our nebula model to further test. It would be an interesting possibility to determine the column density of the dusty H I component or even detect a molecular component with JWST/MIRI or ALMA. Moreover, deep Chandra X-ray observations might constrain cluster wind/ejecta radiative cooling and determine whether radiation feedback is more important than kinetic feedback in the system.

\section*{acknowledgments}

{Some of the data used in this paper were obtained from the Mikulski Archive for Space Telescopes (MAST) at the Space Telescope Science Institute. The specific observations analyzed can be accessed via this \dataset[DOI]{https://doi.org/10.17909/tznp-gv14}}.

The authors would like to thank the anonymous referee, whose suggestions improved this manuscript. The authors would also like to thank Xiao Fang, Brenda Frye, Alex Ji, Wenbin Lu and Chris Matzner for helpful discussions. MP was funded through the NSF Graduate Research Fellowship grant No.~DGE 1752814, and acknowledges the support of System76 for providing computer equipment. LD acknowledges research grant support from the Alfred P. Sloan Foundation (Award Number FG-2021-16495).
The research of BT is funded by the Gordon and Betty Moore Foundation through Grant GBMF5076, and by NASA ATP-80NSSC18K0560 and ATP-80NSSC22K0725.
This research was supported in part by the National Science Foundation under Grant No. NSF PHY-1748958.
The research of CFM is supported in part by NASA ATP grant 80NSSC20K0530.



\bibliography{lyc}{}
\bibliographystyle{aasjournal}



\end{document}